\documentclass[longauth]{aa}

\usepackage{soul}
\usepackage{graphicx}
\usepackage{color}
\usepackage{txfonts}
\usepackage{natbib}
\usepackage[squaren,Gray]{SIunits}

\usepackage{bbold}
\usepackage{subcaption}
\usepackage{mwe}

\newcommand{\sink}{\mathrm{sink}}

\renewcommand{\vec}[1]{\mathbf{#1}}

\usepackage[colorlinks=true,citecolor=blue]{hyperref}
\usepackage{orcidlink}

\begin{document}

   \title{The Rosetta Stone Project}

   \subtitle{I. A suite of radiative magnetohydrodynamics simulations of high-mass star-forming clumps}

      \titlerunning{Rosetta Stone. Paper I}

   \author{          Ugo~Lebreuilly\inst{\ref{Saclay}}\orcidlink{0000-0001-8060-1890}
\thanks{ugo.lebreuilly@cea.fr}   
   \and
Alessio~Traficante\inst{\ref{iaps}}\orcidlink{0000-0003-1665-6402}
   \and
Alice~Nucara\inst{\ref{iaps}}\fnmsep\inst{\ref{ToV}}\orcidlink{0009-0005-9192-5491}
   \and
   Ngo-Duy~Tung\inst{\ref{Saclay}}\orcidlink{0009-0009-8545-2682}
   \and
   Patrick~ Hennebelle\inst{\ref{Saclay}}\orcidlink{0000-0002-0472-7202}
  \and
     Sergio~Molinari\inst{\ref{iaps}}\orcidlink{0000-0002-9826-7525}
   \and Ralf~S.~Klessen\inst{\ref{Heidelberg1}}\fnmsep\inst{\ref{Heidelberg2}}\fnmsep\inst{\ref{Cfa}}\fnmsep\inst{\ref{Radcliffe}}\orcidlink{0000-0002-0560-3172}
   \and  Leonardo~Testi\inst{\ref{DIFA}}\fnmsep\inst{\ref{Arcetri}}\orcidlink{0000-0003-1859-3070}
   \and  Veli-Matti~Pelkonen\inst{\ref{iaps}}\orcidlink{0000-0002-8898-1047}
   \and Milena~Benedettini\inst{\ref{iaps}}\orcidlink{0000-0002-3597-7263}
   \and   Alessandro~Coletta\inst{\ref{iaps}}\orcidlink{0000-0001-8239-8304}
   \and  Davide~Elia\inst{\ref{iaps}}\orcidlink{0000-0002-9120-5890}
   \and Chiara~Mininni\inst{\ref{iaps}}\orcidlink{0000-0002-2974-4703}
       \and  Stefania~Pezzuto\inst{\ref{iaps}}\orcidlink{0000-0001-7852-1971}
   \and Juan~D.~Soler\inst{\ref{iaps}}\orcidlink{0000-0002-0294-4465}
   \and 
 Paolo~Suin\inst{\ref{Marseille}}\orcidlink{0000-0001-7044-3809}
 \and
Claudia~Toci\inst{\ref{Arcetri}}\fnmsep\inst{\ref{ESO}}\orcidlink{0000-0002-6958-4986}
}

\institute{Université Paris-Saclay, Université Paris Cité, CEA, CNRS, AIM, F-91191, Gif-sur-Yvette, France \label{Saclay}
\and 
INAF - Istituto di Astrofisica e Planetologia Spaziali, Via Fosso del Cavaliere 100, I-00133 Roma, Italy \label{iaps}
\and
Dipartimento di Fisica, Università di Roma Tor Vergata, Via della Ricerca Scientifica 1, I-00133 Roma, Italy \label{ToV}
         \and
         Universität Heidelberg, Zentrum für Astronomie, Institut für Theoretische Astrophysik, Albert-Ueberle-Str. 2, 69120 Heidelberg, Germany \label{Heidelberg1}
                \and 
         Universität Heidelberg, Interdisziplinäres Zentrum für Wissenschaftliches Rechnen, Im Neuenheimer Feld 205, 69120 Heidelberg, Germany \label{Heidelberg2}
         \and
         Harvard-Smithsonian Center for Astrophysics, 60 Garden Street, Cambridge, MA 02138, U.S.A. \label{Cfa}
         \and
         Radcliffe Institute for Advanced Studies at Harvard University, 10 Garden Street, Cambridge, MA 02138, U.S.A. \label{Radcliffe}
                  \and
                           Alma Mater Studiorum Università di Bologna, Dipartimento di Fisica e Astronomia (DIFA), Via Gobetti 93/2, I-40129, Bologna, Italy \label{DIFA}
                        \and
         INAF-Osservatorio Astrofisico di Arcetri, Largo E. Fermi 5, I-50125, Firenze, Italy \label{Arcetri}
                 \and
         Aix Marseille Univ, CNRS, CNES, LAM Marseille, France \label{Marseille}
                  \and
         European Southern Observatory (ESO), Karl-Schwarzschild-Strasse 2, 85748, Garching bei Munchen, Germany \label{ESO}
         }
        
   \date{Received month dd, yyyy; accepted month dd, yyyy}

  \abstract{Star formation and, in particular, high-mass star formation are key astrophysical processes that are far from being fully understood. Unfortunately, progress in these fields is slow because observations are hard to interpret as they cannot be directly compared to numerical simulations. Synthetic observations are therefore necessary to better constrain the models.}{With the Rosetta Stone project, we aim to develop an end-to-end pipeline to compare star formation simulations with observations as accurately as possible in order to study the evolution from clumps scales to stars.}{Using the adaptive mesh-refinement code RAMSES, we computed a first grid of model of star-forming clumps to develop our pipeline and explore the impact of the clump initial conditions on their evolution. The main purpose of this set of simulations is to be converted into synthetic observations to enable a direct comparison with real star-forming clumps observed with Herschel and ALMA. }{The Rosetta Stone simulations presented here provide a catalog available for full post-processing and subsequent comparison with observations (RS1). Among all the parameters explored here, the strength of the magnetic field has the strongest influence on the clump evolution (fragmentation, star formation, global collapse) at both large and small scales. Numerical parameters such as the resolution per Jeans length or the threshold for accretion onto sink particles affects the formation of low-mass sinks. Finally, the widely used $L/M$ ratio is found to be a good indicator of the clump evolutionary state regardless of its initial condition, but this could change when more feedback processes (jets, HII regions) are included.}{We now have a new suite of simulations of star-forming clumps that is available for full post-processing and subsequent comparison with the observations. }
      \authorrunning{Lebreuilly et al}
  \maketitle
  
\section{Introduction}

Massive stars are among the most fundamental astrophysical objects. They are
primary agents of the chemical and dynamical evolution of galaxies.
They inject vast amounts of energy into their environments, during their birth through energetic jets and at their death through supernova explosions \citep[see, for example,][]{krumholz2014,2016ARA&A..54..491B}.
Throughout their lifetimes, they also introduce strong UV radiation and winds, which drive the expansion of ionized (H{\sc ii}) regions and potentially play a critical role in the dispersal of molecular clouds and the regulation of star formation \citep{2016SAAS...43...85K}. 

Despite their crucial role, it is still unclear exactly how massive stars form \citep[see the reviews by][]{2007ARA&A..45..481Z,2018ARA&A..56...41M}. Traditionally, the community has explored two families of models; namely, core-fed (core accretion) and clump-fed (competitive accretion). In the first scenario, massive stars are born from massive prestellar cores, i.e., a single mass reservoir \citep[see for e.g.,][]{2003ApJ...585..850M,2014prpl.conf..149T}. In the clump-fed scenario, which seems to be well supported by observations (\citealt{Anderson2021};\citealt{SQUALO}; \citealt{2025arXiv250305663C} ), no massive prestellar core is required; massive stars are the result of the hierarchical collapse of massive star-forming clumps and protostars can get their mass from matter that is not part of their initial reservoir \citep{2004MNRAS.349..735B,2019MNRAS.490.3061V}. Recently, some simulations of star-forming clumps \citep{2018A&A...611A..24H} and Herschel observations of NGC 6334 and Aquila \citep{2021A&A...653A.157L} have complicated this picture, sowing questions about the reality of cores as bound objects even at lower masses. This hints that star formation could be regulated at even smaller scales by the change of opacity at high density, which controls the formation of the first Larson core \citep{2019ApJ...883..140H,2023arXiv230816268G}.

 To make progress with these questions, observations are, of course, key tools for astrophysicists.
Observations indicate that massive stars generally do not form in isolation but rather in massive complexes, called protostellar clumps, which are the dense regions of molecular clouds \citep{Ladalada,2010ARA&A..48..431P}. Infrared and millimeter observations of nearby and more distant star-forming regions allow us to explore how these star-forming environments fragment from clumps to cores, and eventually to stars (\citealt{2017MNRAS.471..100E};  \citealt{Urquhart2014}; \citealt{2017MNRAS.470.3882T}; \citealt{Urquhart2018}; \citealt{2019ApJ...886..102S}; \citealt{ALMA-IMF}; \citealt{2024ApJS..270....9X}; \citealt{2025A&A...696A.149M}). However, interpreting observations is generally extremely difficult. First of all, the finite resolution (beam size) and sensitivity of the instruments generally makes us blind to small physical scales despite the important role they are bound to play. Second, it is rarely possible to reproduce the 3D structures of observations and we must therefore interpret projections along the line of sight. In addition, we rely on various models such as dust models, radiative transfer models, SED models for the stars to interpret these observations. Also, some crucial quantities, known to impact the dynamics of such environment (magnetic field, ionization rates, dust size distribution), are generally extremely hard to measure with confidence from the observations. Furthermore, it is impossible to access the time evolution on the timescales of the star formation mechanisms, typically a few $10^5$ years \citep{2018ARA&A..56...41M}. Therefore, we need to use statistics to compensate. Last but not least, observations are as complex
as they can be since all these physical ingredients are present at the same time, interacting in a complex, nonlinear way. 

It is therefore extremely important to also resort to numerical simulations as a means of interpreting the observations. While they necessarily approximate reality, they allow us to isolate the role of crucial parameters and the initial conditions, and give us direct access to the physical conditions in the object of interest (density, velocity, temperature, energy budget, time evolution, etc.). In the star formation community, a wide class of models going from galactic scales down to kiloparsecs, then to clump scales and eventually individual star-forming events have been proposed. All of them are of course useful in their own way, exploring different scientific questions such as the star formation rate in galaxies \citep[e.g.,][]{2012ApJ...761..156F,2018A&A...620A..21C,2019ApJ...879..129B,Brucy2020}, the initial mass function (IMF) of the stars \citep[e.g.,][]{LeeandHennebelle2018,2020MNRAS.492.4727C,2021MNRAS.503.1138N,2021MNRAS.502.3646G,Hennebelleetal2022,2023MNRAS.518.5190M}, the formation of low- \citep[e.g.,][]{2013ApJ...763....6T,2018A&A...615A...5V,2020A&A...638A..86B,2023A&A...680A..23A,2023ApJ...946....3K} and high-  \citep[e.g.][]{2022A&A...658A..52C,2022ApJ...941..202R,2023A&A...669A..80O} mass stellar systems, or even trying to get a complete picture of star-forming environments from intermediate \citep{Girichidis2011,Girichidis2012a,Girichidis2012b,2017MNRAS.472.4797S,2018MNRAS.480.3511G,2017ApJ...846..133K,2018A&A...611A..24H,2021MNRAS.506.2199G,2023A&A...675A.144B} to galactic scales \citep{2013MNRAS.436.1836R, 2020MNRAS.499.4455T,2020MNRAS.492.1594S, 2021MNRAS.503.5826A,2022MNRAS.515.1663J,2023A&A...672A.193F,2024ApJ...974..240Z,2025arXiv250202646G}. Unavoidably, they all suffer some shortcomings. First of all, they also have a finite resolution, i.e., a smallest scale they can actually probe. But, contrary to observations, this limitation affects the simulation results at both small and large scales since inconsistency at small scales cascades up to large scales. Indeed unresolved processes can, in principle, feedback on large scales. It is customary to resolve these effects with sub-grid modeling (such as sink particles for example). Naturally, these models usually rely on many simplifying hypothesis and therefore can be flawed and even sometimes inaccurate. The finite resolution of simulations is also a major problem that generates numerical errors since a discretization in time and space is required to solve the physical equations. As there are various ways to discretize the equations, many numerical codes have been developed to study star formations over the past 20-30 years (with grid-based schemes \citealt{2000ApJS..131..273F}; \citealt{RAMSES}; \citealt{2007ApJS..170..228M}; \citealt{2016ApJS..223...11B}; \citealt{2021JOSS....6.2807P}, smoothed-particle hydrodynamics \citealt{2018PASA...35...31P}, or moving meshes \citealt{2010MNRAS.401..791S}; \citealt{2015MNRAS.450...53H}). Each method has its own advantages and inconveniences, often balancing accuracy (with more or less high-order schemes) and robustness (with more or less numerical diffusivity), and they never perfectly agree with each other \citep[see e.g.][]{2024ApJ...970..156D}. Another problem of numerical models resides in the relatively arbitrary choice of initial conditions \citep[see][for a discussion on the subject]{2022MNRAS.510.4767L}. Those are always idealized in simulations, which leads to a lack of realism and often to large scales not being properly accounted for. While it is possible to start from relatively large-scale initial conditions and zoom in \citep{Kuffmeier2017,Kuffmeier2019,2017MNRAS.472.4797S,2018A&A...611A..24H}, a lot can still be learned from simpler isolated setups in a more controlled environment.

Once simulations are produced, they can be compared with observations after being post-processed through the radiative transfer tools developed in recent years \citep{2006A&A...459..797P,2011ApJS..196...22B,RADMC3D,2016A&A...593A..87R}. These codes allow one to generate ideal sky images of the dust continuum, molecular lines, or nonthermal radiation processes. These images can then be passed to a simulator such as CASA \citep{CASA} to reproduce the sensitivity, sky coverage, and other artifacts brought by telescopes and interferometers. This produces synthetic observations that can be used for comparison with observations, or to test observational techniques to measure physical quantities such as the density, temperature, and magnetic field strength. Already widely used in studies of evolved protoplanetary disks \citep[see for e.g.][among many other studies]{2015MNRAS.453L..73D,2020ApJ...888L...4T,2022A&A...665A..25C}, synthetic observations have also proven to be extremely useful tools to make progress on our understanding of star-forming clumps \citep{2008AJ....136..404O,2020ApJ...900...82P,2022arXiv220607440Z}, forming protostars and their feedback \citep{2012A&A...545A..98C,2014ApJ...784...61O,2018MNRAS.477.2760M,2022A&A...668A..83V}, and nascent protostellar disks \citep{Tung2024} and bridge gaps between the observer and theoretical communities. However, the systematic production of these observations remains a rather rare methodology in the star formation community and more effort should be put into it to get a full picture of the star-forming interstellar medium (ISM)'s physical evolution.

In this spirit, we propose a new framework called The Rosetta Stone project \footnote{\url{https://www.the-rosetta-stone-project.eu}} to compare star-forming clump simulations with real observations by ALMA and Herschel. This project aims to help us better understand the conditions at play in real star-forming clumps and their role in the clumps' subsequent evolution (fragmentation, star formation, dispersion). We devise a first catalog of models (RS1) of high-mass star-forming clumps with various initial masses, magnetic fields, turbulent strengths, initial random turbulent fields, and assumptions about numerical parameters. In this first paper of the series (Paper I), we provide a physical description of our model catalog, We discuss the dynamical, thermal, and magnetic evolution of the clumps as well as the subsequent star formation. We also discuss our parameter choice, the limitation of our models, and our prospects of reaching better fidelity to real star-forming clumps. In our second paper (hereafter Paper II), {\color{blue} Tung et al. (in prep)}, we present our post-processing routines for produced ideal sky images with RADMC-3D \citep{2012ascl.soft02015D} at ALMA and Herschel wavelengths. The conversion into Herschel images will also be presented as well as a study to assess the luminosity-to-mass ratio ($L/M$) as a means of determining the evolutionary state of the clumps. Finally, paper III {\color{blue}(Nucara et al. in prep)} is dedicated in a comparison between this first grid of models and the clump observations of the SQUALO survey \citep{SQUALO} by converting the perfect sky images at ALMA wavelength into synthetic observations with CASA. This comparison will use the same extraction tools as SQUALO to investigate the fragmentation levels of numerical simulations and interpret the SQUALO observations.

 This paper is organized as follows. In Sect.~\ref{sec:methods}, we describe our numerical methods as well as our simulation plan. Then, we present the results of our reference simulations and parameter exploration in Sect.~\ref{sec:results} and Sect.\ref{sec:explore}. In Sect.~\ref{sec:discussion}, we discuss the implications of these results, and we propose our conclusions and prospects in Sect.~\ref{sec:conclusion}.

\section{Methods}
\label{sec:methods}

\subsection{Physical equations}

We investigated the evolution of star-forming clumps by solving the equation of radiative magnetohydrodynamics (RMHD) with self-gravity, which reads as follows:
\begin{eqnarray}
\frac{\partial \rho}{\partial t} &+&\nabla \cdot \left[ \rho \vec{v} \right] = \nonumber 0,\\
\frac{\partial \rho \vec{v}}{\partial t}  &+& \nabla \cdot \left[ \rho \vec{v} \vec{v} + (P_{\rm{th}} + \frac{\vec{B}^2}{2}) \mathbb{I} -\vec{B} \vec{B} \right] = - \rho  \vec{\nabla} \phi -\lambda \nabla E_{\rm{r}}, \nonumber \\
\frac{\partial E}{\partial t}  &+&\nabla \cdot \left[\vec{v} (P_{\rm{th}} +E +\frac{\vec{B}^2}{2}) - \vec{B (\vec{B} \cdot \vec{v)}}\right]= - \rho \vec{v} \cdot \vec{\nabla} \phi   \nonumber \\  &-&\mathbb{P}_{\rm{r}} \nabla{:} \vec{v} -\lambda \vec{v} \nabla \cdot E_{\rm{r}}   \nonumber \\ 
&+& \nabla \cdot \left(\frac{c \lambda}{\rho \kappa_{\rm{R}}} \nabla E_{\rm{r}} \right) +  S_{\star} ,\nonumber \\
\frac{\partial E_{\rm{r}}}{\partial t}  &+&\nabla \cdot \left[\vec{v} E_{\rm{r}} \right] = - \mathbb{P}_{\rm{r}} \nabla{:} \vec{v}  + \nabla \cdot \left(\frac{c \lambda}{\rho \kappa_{\rm{R}}} \nabla E_{\rm{r}} \right) 
\nonumber \\
&+& \kappa_{\rm{P}} \rho c (a_{\rm{R}} T^4- E_{\rm{r}}) +  S_{\star} , \nonumber \\
\frac{\partial \vec{B}}{\partial t}  &-& \nabla \times \left[\vec{v} \times \vec{B} \right] = 0, \nonumber \\
\nabla \cdot \vec{B}&=&0, \nonumber \\
\triangle \phi &=& 4 \pi \mathcal{G} \rho.
\end{eqnarray}

Here $\rho$, $\vec{v}$, and $E$ are the gas density, velocity, and total energy, respectively. In addition, we defined the thermal pressure, $P_{\rm{th}}$, temperature, $T$, magnetic field, $\vec{B}$, and gravitational potential, $\phi$, as well as the radiative energy, $E_{\rm{r}}$, and pressure, $ \mathbb{P}_{\rm{r}}$. The treatment of the radiative transfer also required us to define the Rosseland and Planck opacities, $\kappa_{\rm{R}}$ and $ \kappa_{\rm{P}}$, the radiative flux limiter, $\lambda$, and the source term from stellar radiation, $ S_{\star} $, which includes the internal and accretion luminosity. Finally, we also defined the gravitational and Stefan-Boltzmann constants, $\mathcal{G}$ and  $a_{\rm{R}}$, as well as the speed of light, $c$.

To solve these dynamical equations, we employed the widely used adaptive-mesh refinement (AMR) code {\ttfamily{RAMSES}} \citep{RAMSES}. In particular, we used its extensions to magnetohydrodynamics \citep{2006A&A...457..371F}, radiative transfer in the flux-limited diffusion (FLD) limit \citep{2011A&A...529A..35C,2014A&A...563A..11C} with stellar feedback \citep{2020ApJ...904..194H} and sink particles \citep{2014MNRAS.445.4015B}.

\subsection{Setup}

We initialized our clumps as uniform spheres of mass, $M_0$ (either 500 or 1000~$M_{\odot}$), with a radius of $R_0=0.38~$pc and a temperature of  $T_0=10$~K. We also assumed a mean molecular weight, $\mu_{\mathrm{g}}=2.31$. At this stage, it is useful to recall the thermal-to-gravitational energy ratio,
\begin{equation}
\alpha \equiv \frac{5}{2} \frac{R_0 k_{\rm{B}} T_{0}}{\mathcal{G} M_0\mu_{\rm{g}}m_{\rm{H}}},
\end{equation}
where $ k_{\rm{B}}$ is the Boltzmann constant and $m_{\rm{H}}$ is the mass of an hydrogen atom. This quantity allows us to verify that the clumps will collapse gravitationally (if $\alpha <0.5$). For a clump of radius 0.38 pc and 1000 (resp. 500) $M_{\odot}$, we have $\alpha = 0.008$ (resp. $\alpha = 0.016$), meaning that they are all gravitationally unstable.

We also threaded the clumps with an initially uniform magnetic field (whose axis defines the z direction) by imposing an initial mass-to-flux over critical mass-to-flux ratio,

\begin{equation}
    \mu = \left(\frac{M_0}{\phi}\right)/\left(\frac{M}{\phi}\right)_c,
\end{equation}
where $\phi = \iint \vec{B} \vec{dS}$ is the magnetic flux through the clump and $\left(\frac{M}{\phi}\right)_c$ is the critical mass-to-flux ratio below which the clump is magnetically supported against the gravitational collapse. In this study, we consider three values for the mass-to-flux ratio: 3 (strong magnetic field), 10 (moderate magnetic field), and 100 (weak magnetic field). As our maximal resolution for this work is about 38 au, we do not consider any nonideal MHD effect. These effects are expected to impact the dynamics of the system only at smaller scales \citep[e.g.,][]{2019MNRAS.489.1719W,Lebreuilly2023,Lebreuilly2024a}.

Finally, we imposed an initial supersonic turbulent velocity field with random phases and following a Kolmogorov power-spectrum of $ \propto k^{-11/3}$.  We point out that, as was shown in the literature, supersonic turbulence follows a power spectrum closer to a Burgers cascade than a Kolmogorov cascade \citep[see for instance][]{2021NatAs...5..365F}. As we are modeling decaying turbulence here, this does not matter significantly. \citep{LeeandHennebelle2018} have shown, in calculations similar to ours, that this choice is not very influential. Indeed the initial conditions of the velocity field are quickly forgotten as the clump proceeds to collapse. A better way to set the velocity field properly would be to zoom in self-consistently from large-scale conditions \citep[as in for e.g.][]{2018A&A...611A..24H}, but this is beyond the scope of the present article. Our current choice of initial condition was made for legacy purposes, as it enables a more direct comparison with existing RAMSES simulations (which usually use seed 1 of the present paper).  In this study, we look at two values for the Mach number, $7$ and $10$. These values, as well as those of the clump mass and size, were chosen to reproduce the turbulence levels of the clumps of the SQUALO survey.

\subsection{Resolution}

We employed the AMR capability of RAMSES to follow the dynamical evolution of the clumps up to a maximal resolution of 38 au. This moderate maximum resolution was a compromise in order to enable us to run a survey of models and was complemented by previous computations from our team made at a resolution of $\sim 1.2~$au \citep{Lebreuilly2024a}. With current super-computing capabilities it is impossible to uniformly resolve the clumps at 38 au resolution; therefore, we used AMR to refine the grid imposing (at least) $N_{\mathrm{jeans}}$ cells per Jeans length. According to the \cite{1997ApJ...489L.179T} criterion, a minimum of eight points are needed to prevent artificial fragmentation. In this work, we consider  $N_{\mathrm{jeans}}=10$ for the survey of models, but we have also explored $N_{\mathrm{jeans}}=20$ for our reference set of parameters.

\subsection{Sink particles}

With our spatial resolution, we do not resolve individual stars. We thus employed sink particles \citep{1995MNRAS.277..362B,2010ApJ...713..269F} to mimic their feedback on the clump evolution. Specifically, we used the RAMSES implementation by \cite{2014MNRAS.445.4015B}. We formed these sinks when the local density reached the density threshold, $n_{\mathrm{sink}}$, for any cell that was not already inside the accretion sphere of an already existing neighboring sink (of radius $4 \Delta x$). Once a sink was formed, it accreted the material with $n>n_{\mathrm{sink}}$ in its accretion sphere at each timestep. 

Accretion triggered a feedback effect as we considered the accretion luminosity of the stars in the energy equation. This accretion luminosity reads as

\begin{equation}
    L_{\rm{acc}} = f_{\rm{acc}} \frac{\mathcal{G} M_{\sink} \dot{M}_{\sink}}{R_{\star}}.
\end{equation}

The factor, $f_{\rm{acc}}<1$, is essentially unknown and depends on how the gravitational energy is converted into radiation at very small scales that are far from being resolved here. In the RS1.0 catalog, we consider $ f_{\rm{acc}}=0.1$ \citep[a value of 0.5 has also been explored in][]{Lebreuilly2024a}. These values close to unity are in lines with the finding of extremely high-resolution simulations of protostar formation \citep{Ahmad2024}. 

The sink internal luminosity, which was computed using the evolutionary tracks of \cite{2013ApJ...772...61K}, has also been included in this work. We point out that those tracks are  also used to compute the stellar radii, $R_{\star}$. This luminosity source is generally much lower than the accretion luminosity at early stages for low-mass stars but is dominant for more massive stars.

\subsection{RS1.0 catalog}

\begin{table}
  \caption{Parameter range of the 24 RS1.0 catalog.}      
\label{tab:catalogue}      
\centering          
\begin{tabular}{c c }     
\hline\hline 
Parameters & Values \\
\hline\hline 
Mass [$M_{\odot}$] & [500,{\bf 1000}] \\
Radius [pc] & [{\bf 0.38}] \\
Mach & [{\bf 7}, 10] \\
$\mu$ & [3, {\bf 10}, 100] \\
Seed & [{\bf{1}} , 2] \\
\hline \hline
\end{tabular}
\tablefoot{ The bold value designed the parameter choice for our reference model.}
\end{table}

\begin{table}
  \caption{Additional individual runs explored for this study.  }      
\label{tab:catalogue2}      
\centering          
\begin{tabular}{ c c c }     
\hline\hline 
 Seed  & $N_{\mathrm{jeans}}$ & $n_{\mathrm{sink}}~[\centi\meter^{-3}]$ \\
\hline\hline 
 3 & 10 & $10^9$ \\
 4 & 10 & $10^9$\\
 1 & 20 & $10^9$ \\
 1 & 10 & $10^{10}$ \\
\hline \hline
\end{tabular}
\tablefoot{These runs were all computed with a clump mass of 1000$M_{\odot}$, radius of 0.38 pc, Mach number of 7, and mass-to-flux ratio of 10.}
\end{table}

\begin{figure*}[h!]
  \centering
 \includegraphics[width=
          0.9\textwidth]{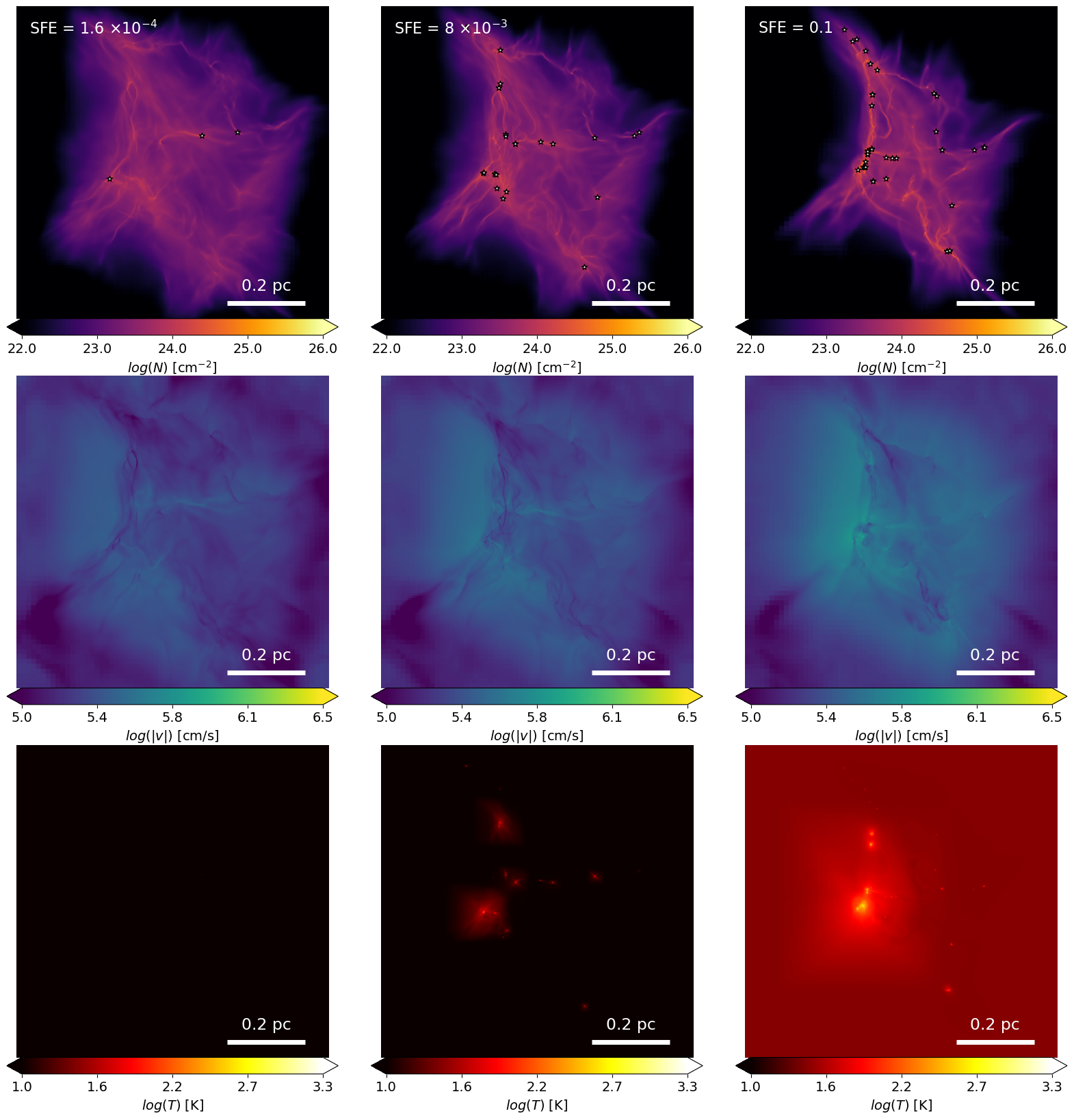}
      \caption{Integrated maps for the reference model in the $z$ direction. From left to right, we display the models at SFE  \,$=$\,$1.6\,\times\,10^{-4}$, SFE\,$=$\,$8\,\times\,10^{-3}$, and SFE=0.1. 
      From top to bottom, we show the column density, mass-averaged velocity, and mass-averaged temperature.
      The open stars in the top panel indicate the projected sink location.
      }\label{fig:coldens_fid}
\end{figure*}

In this section, we present the first survey of 24 models that we computed, as well as some additional models. Table~\ref{tab:catalogue} shows the range of parameters fully 
explored in this model catalog. The parameters 
we chose for our reference model are highlighted in bold. By convention, we have selected our reference model to be defined by the following combination of parameters: a clump mass of 1000\,$M_{\odot}$, a Mach number of 7, a mass-to-flux ratio of 10, seed 1, and a clump radius of 0.38\,pc. This is the same reference set of parameters as in \cite{Lebreuilly2024a}. For the whole RS1.0 catalog, we considered $n_{\mathrm{sink}} =10^{9} \centi\meter^{-3}$ and 10 points per Jeans length. In addition to the RS1.0 models, we ran four additional models (which we label as RS1.1 models) to explore specific parameters. The parameter choice of these models is described in Table~\ref{tab:catalogue2}.

\subsection{Some useful metrics}

Writing the total sink mass as $M_{\mathrm{tot}} = \sum_{\mathrm{i=1}}^{N_{\mathrm{sink}}} M_{\mathrm{i}} $ allows us to define the star (or sink) formation efficiency (or SFE) as 
\begin{eqnarray}
    \mathrm{SFE} \equiv \frac{M_{\mathrm{tot}}}{M_0}.
\end{eqnarray}
This quantity is useful for describing the evolutionary state of the clump and comparing models. It is the measure of the relative conversion of the clump material into stars or sinks. We also defined the star formation rate, $\dot{M}_{\mathrm{tot}}$, as the sum of the individual accretion rate onto each sink. This allows us to describe how fast star formation is occurring, as a whole, within the clump.

Some other sink quantities are very helpful to describe the clump evolution. We defined the median sink mass, $M_{\mathrm{sink,med}}$. This quantity is useful for assessing the position of the peak of the stellar IMF. We also defined the maximal sink, $M_{\mathrm{sink,max}}$; that is, the mass of the most massive star in the clump. This quantity allows us to assess whether the clump forms at least one massive star. 

Defining the total stellar luminosity, $L_{\mathrm{tot}} \equiv \sum_{\mathrm{i=1}}^{N_{\mathrm{sink}}}L_{\mathrm{acc,i}}+ \sum_{\mathrm{i=1}}^{N_{\mathrm{sink}}} L_{\mathrm{int,i}}$, we finally introduced the $L/M$ ratio as 
\begin{eqnarray}
    L/M \equiv \frac{L_{\mathrm{tot}}}{M_0-M_{\mathrm{sinks}}}.
\end{eqnarray}

This quantity measures the relative conversion of the clump material into stars or sinks. It should increase  with time, and thus is widely used as a clock to describe the evolutionary state of clumps \citep{2015MNRAS.451.3089T,2016ApJ...826L...8M,2018MNRAS.477.2220T,2021MNRAS.504.2742E}. We note a difference between our definition and the one of \citet{2021MNRAS.502.3646G}. They indeed use the stellar mass in the denominator rather than the available gas mass (in our case). We chose this definition since it is closer to the observation definition for future comparison (in Paper II).

\section{Reference model}
\label{sec:results}

\begin{figure*}[t!]
  \centering
 \includegraphics[width=0.49
          \textwidth]{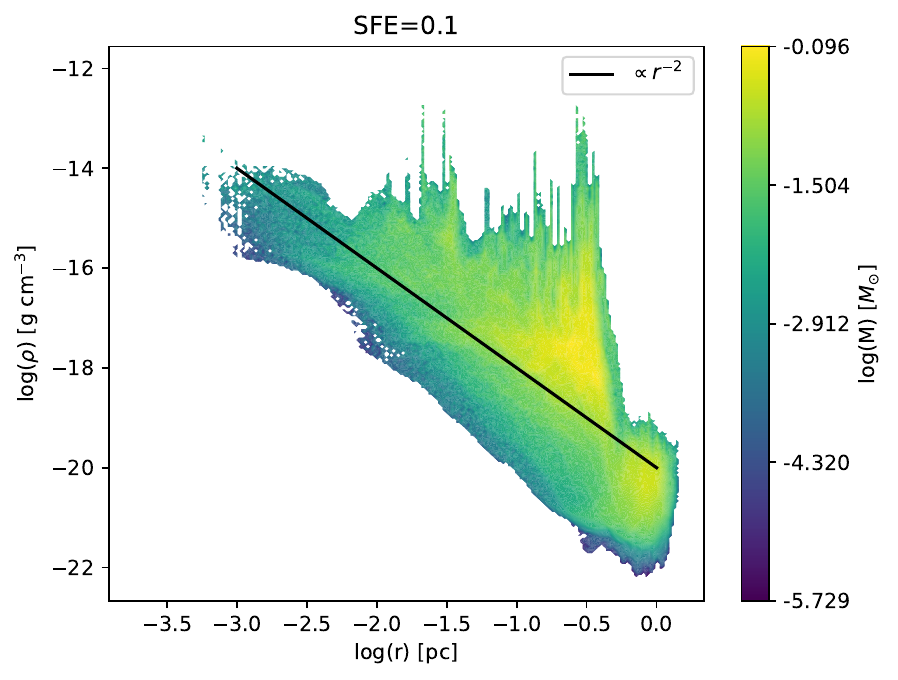}
 \includegraphics[width=0.49
          \textwidth]{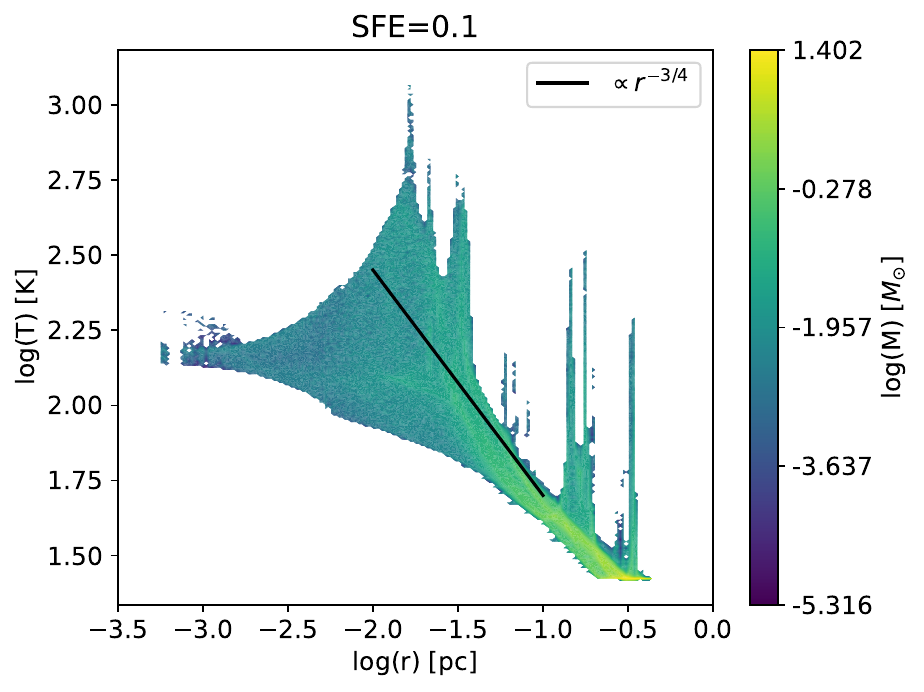}
      \caption{Distribution of physical properties in the reference model.
      {\it Left:} Histogram of the density and radius (centered around the density peak).  We overlay a $r^{-2}$ profile, which is expected in a globally collapsing clump.
      {\it Right:} Histogram of the temperature as a function of the radius (centered around the density peak).  
     The $r^{-3/4}$ profile, shown in black,
      indicates the trend for a star-forming clump at radiative equilibrium for which the temperature is controlled by the stellar luminosity. 
      For both plots, the color map displays the mass in each histogram bin.}
            \label{fig:density_profile}
\end{figure*}
\begin{figure*}[h!]
  \centering
 \includegraphics[width=0.49
          \textwidth]{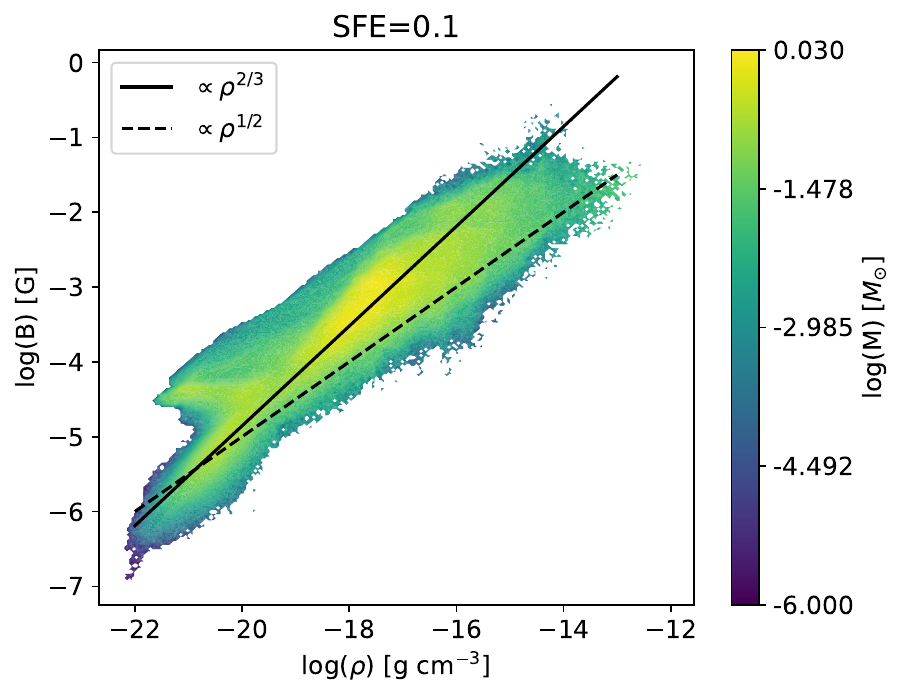}
     \includegraphics[width=0.49
          \textwidth]{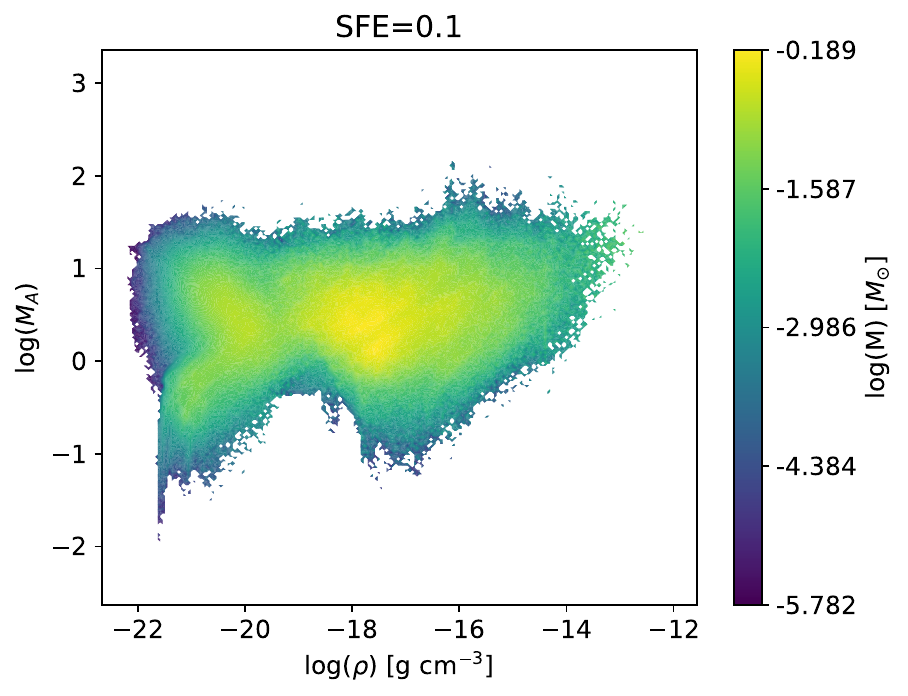}
      \caption{Magnetic field in the reference model. On the left we display the histogram of the magnetic field strength as a function the density. On the right we display the histogram of the Alfv\'enic Mach number vs. the density.  The colors display the mass in each bin of the histogram. The solid and dashed black lines denote a $\propto \rho^{3/2}$ and $\propto \sqrt{\rho}$ profile, respectively. The histograms are consistent with a clump for which the bulk of the collapse is weakly influenced by the magnetic field ($\propto \rho^{3/2}$) and some of the low-density material (with $M_A$\,$<$\,$1$) is supported, in part, by the magnetic field.}
            \label{fig:B_profile}
\end{figure*}

We first present in detail our reference model, which was computed assuming a clump mass of 1000 $M_{\odot}$, a mass-to-flux ratio of 10, and a Mach number of 7. This model was computed with seed 1. As for all the models of RS1.0, we considered $N_{\mathrm{jeans}}=10$, $N_{\mathrm{sink}}=10^9~\centi\meter^{-3}$ and a radius of 0.38 pc. This fiducial set of parameters serves as a reference for comparing the models. 

\subsection{General description of the clump}

We start with a general description of our reference model morphology. In Fig~\ref{fig:coldens_fid}, we present column density (top), mean temperature (middle), and velocity (bottom) maps integrated in the z direction at SFE=$1.6 \times 10^{-4}$, SFE=$8 \times 10^{-3}$, and SFE=0.1. We also display the position of the sink particles in the column density maps. As was expected, the clump undergoes global collapse. This can be seen in both the column density and velocity maps of Fig ~\ref{fig:coldens_fid}. The left panel of Figure ~\ref{fig:density_profile} shows a density histogram displayed as a function of the distance from the density maxima sill at SFE=0.1. This is additional evidence of the global collapse as we can see that the profile is in rough agreement with a $\rho \propto r^{-2}$ profile (with a spread due to the multiple sites of star formation). This scaling is the one expected for a spherically symmetric collapse before the formation of a central object \citep[see the table II and III of][]{2004RvMP...76..125M}. We note that the profile is more complex than this, with the presence of a number of peaks. These are caused by local collapses, which lead to the formation of the individual stars of the simulations. The agreement with the $\rho \propto r^{-2}$ profile is also imperfect because of the turbulence and magnetic field that breaks the symmetry of the collapse \citep[see e.g.][]{2015MNRAS.446.2118T}. We indeed observe a network of filaments and sheets, which is a hallmark of gravo-turbulent motions. Despite clearly not being isotropic, the collapse does trigger the formation of a main hub where most of the star formation activity occurs. This was also shown with previous explorations of the same configuration with higher-resolution models designed to study disk formation \citep{Lebreuilly2024a}. Generally speaking, we note a very close evolution between this model and its higher-resolution counterpart (although protostellar disks are obviously not resolved here) at scales larger than the minimal cell size.

\subsection{Thermal structure}

Let us now turn our attention to thermal properties of the clump at various scales (see the temperature maps in Fig.~\ref{fig:coldens_fid}). While the clump starts at 10 K, star formation quickly leads to significant heating of the regions where the formation is more active. The main star-forming cluster is the hottest region of the
clump, with temperatures reaching about 800 K at the resolution limit. Furthermore, the background temperature also increases and is on the order of 25 K at SFE=0.1, more than two times its initial value. It then reaches $\sim 40~$K at later times, not shown in the figures, as the stars become even more luminous. This background temperature is entirely consistent with the ones estimated in the observations of the Hi-Gal clumps by fitting the spectral energy distribution in five infrared bands \citep[][]{2015MNRAS.451.3089T,elia2017}. We point out that the temperature also varies radially. In the right panel of Fig.~\ref{fig:density_profile}, we show a histogram of the temperature as a function of the distance from the density peak. We clearly see that the temperature has a power law dependency with the radius in the bulk of the mass in the clump. This profile is consistent with the $\propto r^{-3/4}$ dependency that was derived by \cite{Hennebelleetal2022} for a star-forming clump at radiative equilibrium under the influence of stellar radiation. As for the density  histogram, we note the presence of several peaks that coincide with the position of the different sinks (which are the luminosity sources in the model). We point out that, as in \cite{Hennebelleetal2022}, the dominant source of radiation is the accretion luminosity of forming protostars at early stages, while the internal luminosity takes over later on.

\subsection{Magnetic field structure}

Finally, we look at the magnetization of the clump. Information about the magnetic field is particularly important as this quantity is extremely difficult to extract with confidence from the observations despite playing a crucial dynamical role. The left panel of Fig~\ref{fig:B_profile} shows a 2D histogram of the magnetic field strength as a function of the density at SFE=0.1. Here the scaling as a function of the density is not really clear. The bulk of the mass follows the $\propto \rho^{2/3}$ scaling, which is a consequence of flux conservation during the collapse. In addition, some of the material follows a $\propto \sqrt{\rho}$ dependency. This behavior starts to make sense when looking at the histogram of the Alfv\'enic Mach number (Fig~\ref{fig:B_profile}, right panel). We see clearly a large fraction of the clump (in mass) is super-Alfv\'enic, with $1<M_A<100$. As is shown in \cite{2017ApJ...838...40M} who explore clumps with various Alfv\'enic Mach numbers \citep[and initially theorized by][]{1966MNRAS.133..265M}, a $\propto \rho^{2/3}$ scaling is consistent with a spherical collapse for which the action of the magnetic field is negligible. On the contrary, strong magnetic fields influence the dynamics of the collapse, making it  more efficient perpendicular to the field lines. In this condition a $B \propto \rho^{1/2}$ scaling is expected. This is most likely what happens here for regions with low Alfv\'enic Mach numbers. We point out that the scaling we obtain is also consistent with the observations of magnetic fields in dense cores \citep{1999ApJ...520..706C,2010ApJ...725..466C}, which suggest a dependency on the order of  $B \propto \rho^{0.5-0.65}$.

\subsection{Stellar and luminosity distribution}

\begin{figure}[t!]
  \centering
 \includegraphics[width=0.4
          \textwidth]{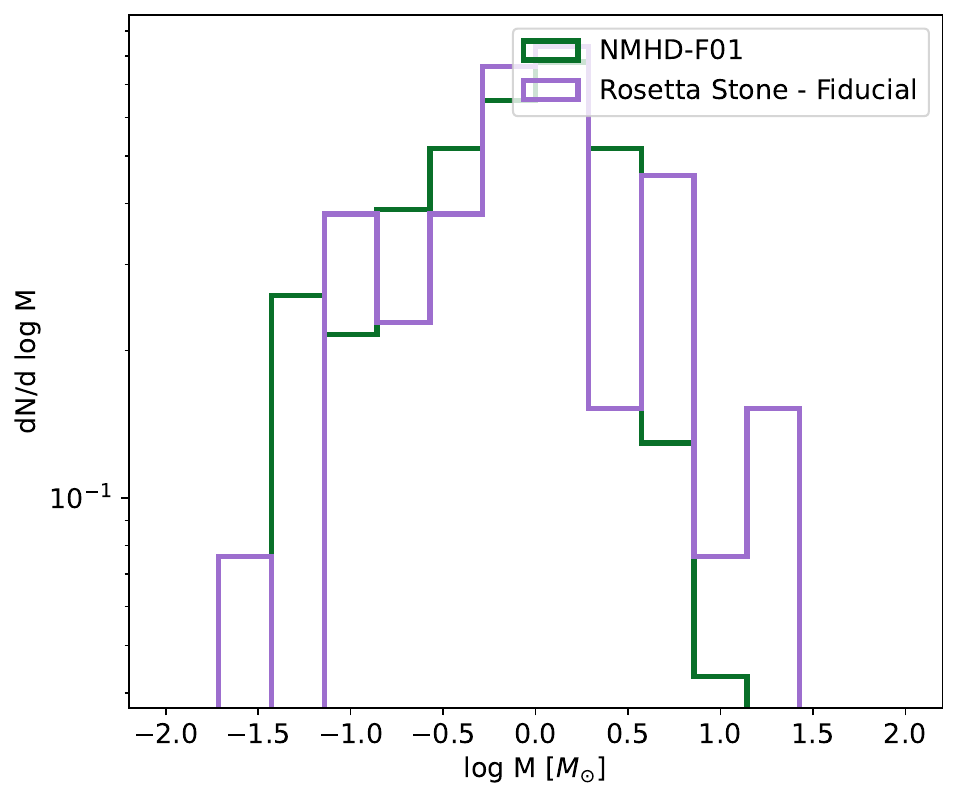}
      \caption{Initial mass function of the sink at SFE=0.1 for the reference model (blue line) and its high-resolution counterpart from \cite{Lebreuilly2024a}. Although some differences can be observed between the two IMFs (in particular at high masses), they roughly agree, which justifies our choice of sink formation threshold.}
            \label{fig:IMF_fid}
\end{figure}

We now describe the stellar IMF, or the sink  mass function to be exact, which we extracted from our reference model.  In Fig~\ref{fig:IMF_fid} we show a comparison between the IMF of our reference model with the higher-resolution counterpart of \cite{Lebreuilly2024a}. The peak of the latter is most likely numerically converged as it resolves the opacity limits \citep{IMFpatrick} and can therefore be used as a reference for the RS model.

First of all, we note a good agreement between the peaks (median) of the two IMFs. This indicates that our choice of $n_{\mathrm{sink}}=10^{9} \centi\meter^{-3}$ is reasonable. Importantly, this peak is located at around $\sim 1 M_{\odot}$. This is quite clearly higher than the peak of the galactic IMF, which is located at around $0.3M_{\odot}$. From these observations, two possibilities appear: either the peak of the IMF is not so universal and we are exploring peculiar conditions with respect to close-by star-forming regions, or some physical effect is missing. While some studies did show that the IMF might no be so universal \citep{Hennebelleetal2022}, \cite{2024A&A...683A..13L} have also shown that protostellar jets make it possible to recover a peak seemingly located at around $0.3M_{\odot}$ for the exact same configuration. Other works also confirm the influence of jets on the IMF \citep{2014ApJ...790..128F,2021MNRAS.502.3646G}.  This justifies a future set of Rosetta Stone catalogs with more complete feedback processes included. If we now turn our attention to low-mass stars ($<0.3M_{\odot}$), we see that the high-resolution run has more of them. This was expected since the adiabatic transition, or opacity limit, which controls the mass of the first Larson core \citep{1969MNRAS.145..271L} and in turn of low-mass stars \citep{2019ApJ...883..140H} is not resolved in the RS simulation. This is one drawback of running the simulations at a lower maximal resolution, which is compensated for by our ability to run full suites of models.

\begin{figure*}[t!]
  \centering
 \includegraphics[width=
          0.8\textwidth]{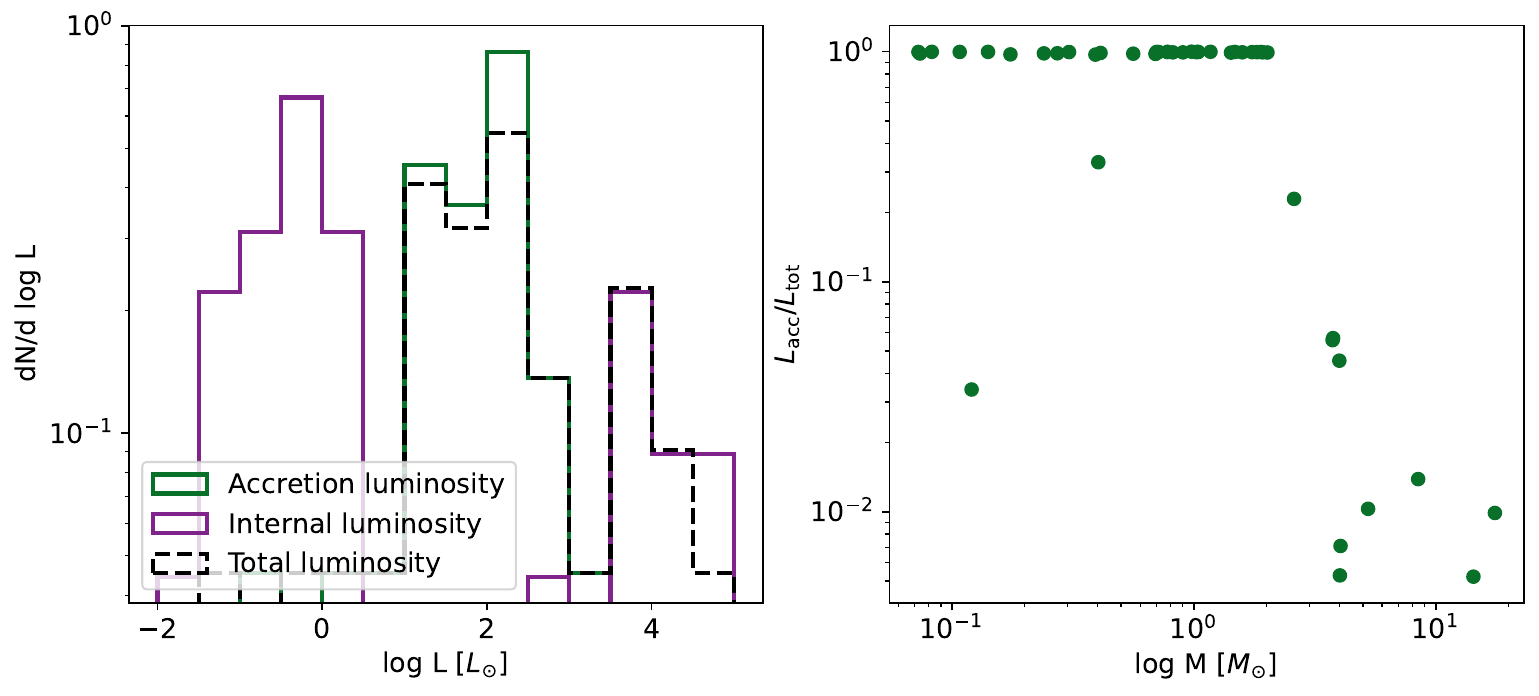}
      \caption{Distribution of accretion, internal and total luminosity (left), and ratio between the accretion and total luminosity as a function of stellar mass (right) at SFE=0.1 for our reference model .}
            \label{fig:Lum_fid}
\end{figure*}

We finish the exploration of our fiducial model by describing the distribution of sink internal and accretion luminosity. Figure~\ref{fig:Lum_fid} shows the distribution of the accretion (green), internal (purple), and total (black) luminosity for our reference model (at SFE = 0.1) as well as the evolution of the ratio $L_{\mathrm{acc}}/L_{\mathrm{tot}}$ as a function of the sink mass (right panel). We see that the accretion luminosity peaks around $100~ L_{\odot}$, whereas the internal stellar luminosity distribution is bimodal, with one peak slightly below  $L_{\odot}$ and another one close to $ 10^4~L_{\odot}$. Looking at the total luminosity distribution, it becomes clear that the accretion luminosity dominates for most objects except for the very bright ones. The low internal luminosity mode therefore applies to most of the stars, i.e., the low-mass ones. The distribution of  $L_{\mathrm{acc}}/L_{\mathrm{tot}}$ clearly shows that the internal luminosity is negligible for stars below $2 M_{\odot}$ and then steeply rises and dominates the total luminosity budget. Only two stars have a low $L_{\mathrm{acc}}/L_{\mathrm{tot}}$ ratio below  $2 M_{\odot}$. These peculiar objects are low accretors and have a low accretion luminosity rather than a high internal luminosity.

Above $2 M_{\odot}$ the internal luminosity steeply rises above the accretion luminosity to reach values on the order of $ \sim 10^4~L_{\odot}$. Of note, at SFE=0.1, the four most massive stars represent 90 \% of the total luminosity and the most massive star represents 50\% of it. The thermal evolution and fragmentation and IMF is influenced by its luminosity source. It is therefore very important to properly predict the high-mass end of the IMF within star-forming clumps. The comparison with observations, which is the goal of the Rosetta Stone project, is of great interest to making progress in the hunt for signs of high-mass star formation. It allows one to directly compare the values of the radiative fluxes from the observer's perspective, link them to the luminosity of the internal sources, and check for potential agreements or disagreements.

\section{Parameter exploration}
\label{sec:explore}
In this section we describe the influence of some key parameters, such as the random initial seed for the turbulence, the magnetization, and the initial Mach number.  As a complement to the results displayed in this section, we show in appendix the evolution of the SFE as a function of time as well as the number of sinks as a function of the SFE for all the RS1.0 models.

\subsection{Random initial conditions}

\begin{figure*}[h!]
  \centering
 \includegraphics[width=
          \textwidth]{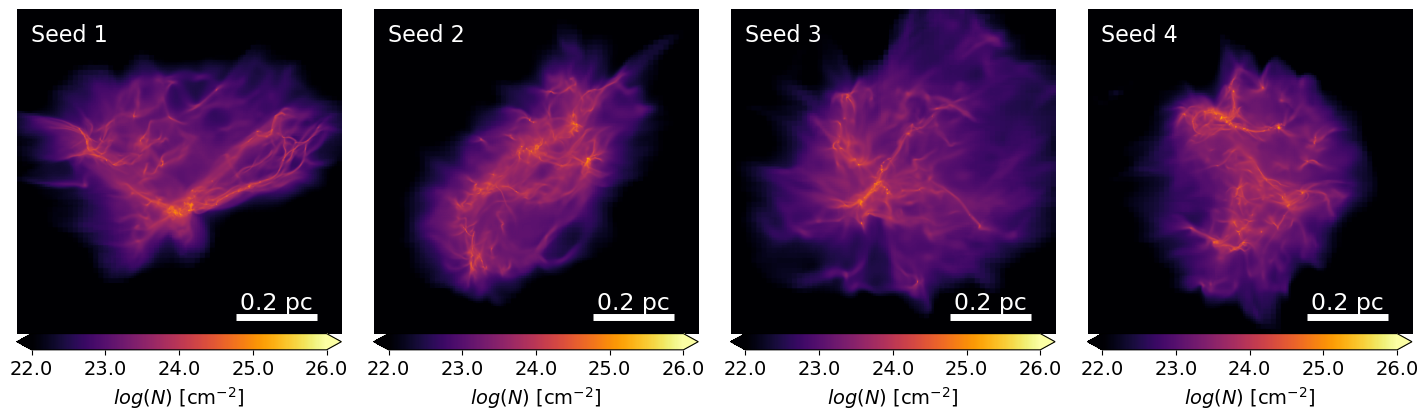}
      \caption{Four realizations of the same star-forming clump with $M=1000~M_{\odot}$, $\mathcal{M}=7$, and $\mu=10$ (from left to right: seed 1, seed 2, seed 3, and seed 4). We display (y-z) maps of the column density integrated in the $x$ direction.}
            \label{fig:seeds_column}
\end{figure*}

\begin{figure}[h!]
  \centering
 \includegraphics[width=0.4\textwidth]{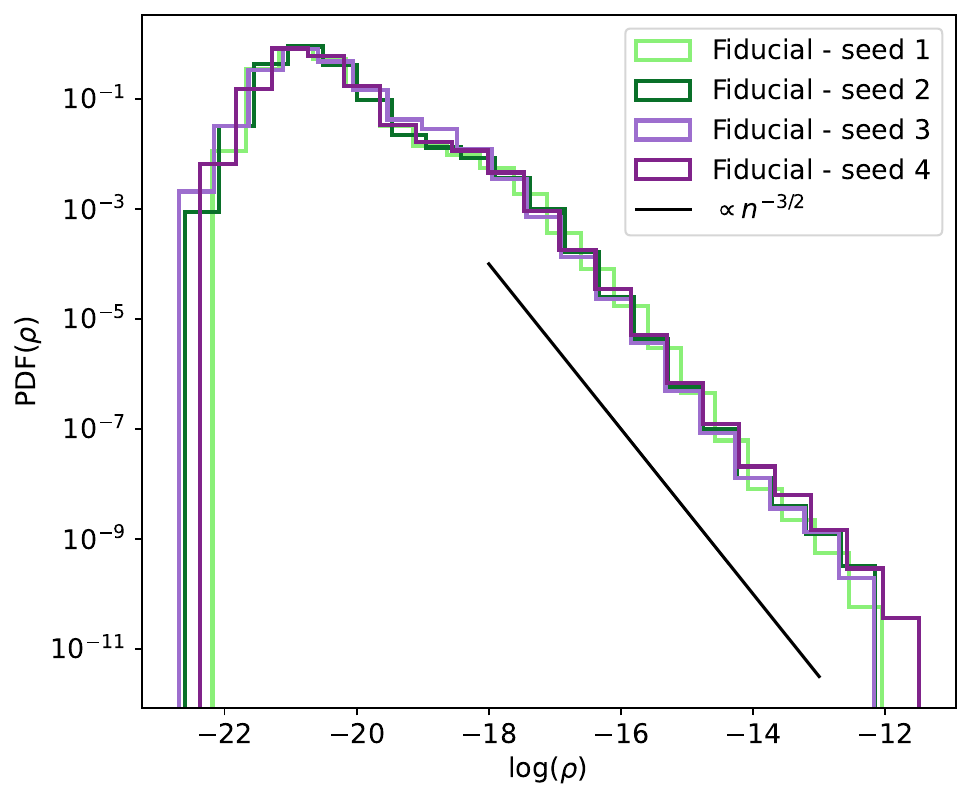}
      \caption{Density PDFs of the four realizations of the clump with $M=1000~M_{\odot}$, $\mathcal{M}=7$, and $\mu=10$. The black line indicates the $\propto n^{-3/2}$ profile expected from  $\propto r^{-2}$ density profile indicative of the collapsing behavior of the clump.}
            \label{fig:PDF_seeds}
\end{figure}

We now discuss the impact of the random initial conditions. We explore four different random seeds (seeds 1, 2, 3, and 4, only the two first being fully investigated in RS1.0), assuming for the clumps that $M=1000 M_{\odot}$, $\mathcal{M}=7$, and $\mu=10$. The column density of the four realizations at SFE=0.1 can be seen in Fig.~\ref{fig:seeds_column}. Without a doubt, the four clumps are visually very different. They nevertheless show many qualitative similarities. They have all fragmented and collapsed in an anisotropic way. Finally, they all form at least one concentrated stellar cluster where most of the star formation appears. The four clumps also show visible filamentous and sheet-like structures, although they are much clearer for seed 1 in the selected projection. Strikingly, a strong similarity in the density PDFs (shown in Fig.~\ref{fig:PDF_seeds}) can be observed. The four PDFs show a $\propto n^{-3/2}$ slope at high density, which is expected of a $r^{-2}$ density profile, and a clear sign of the collapse of the clump taking place.

\begin{figure}[h!]
  \centering
 \includegraphics[width=0.4
          \textwidth]{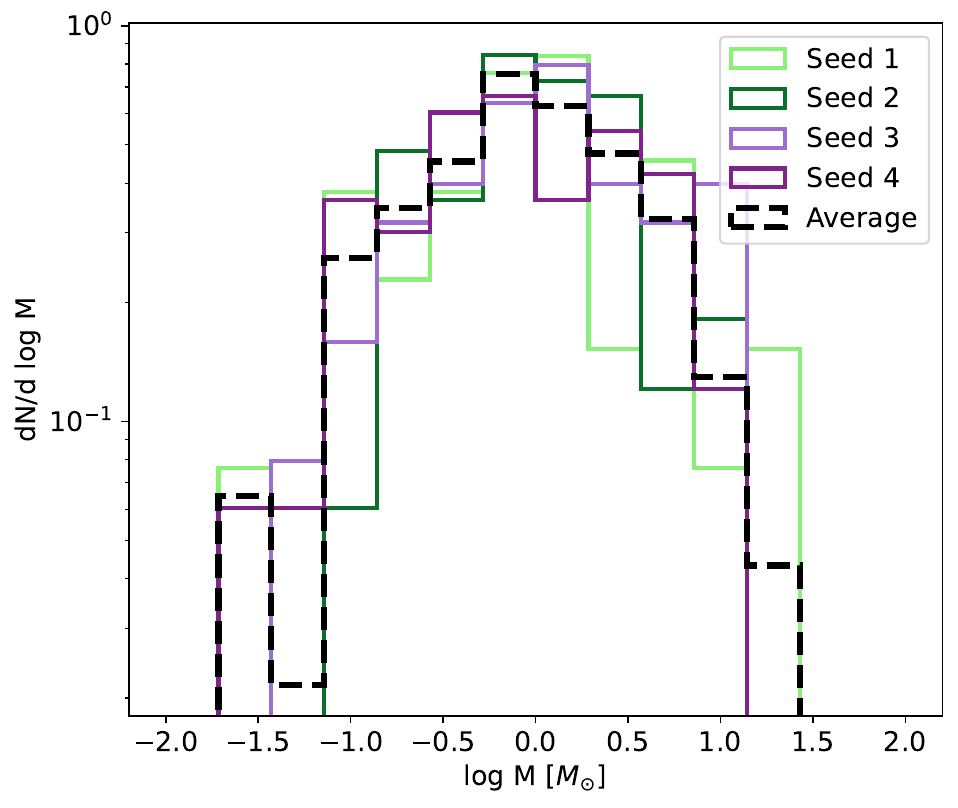}
      \caption{Sink mass function
      at SFE=0.1 for the four realizations of the same star-forming clump with $M=1000 ~M_{\odot}$, $\mathcal{M}=7$, and $\mu=10$. The black line shows the IMF average out of the four clumps.}
            \label{fig:IMF_seeds}
\end{figure}

\begin{figure}[h!]
  \centering
 \includegraphics[width=0.45\textwidth]{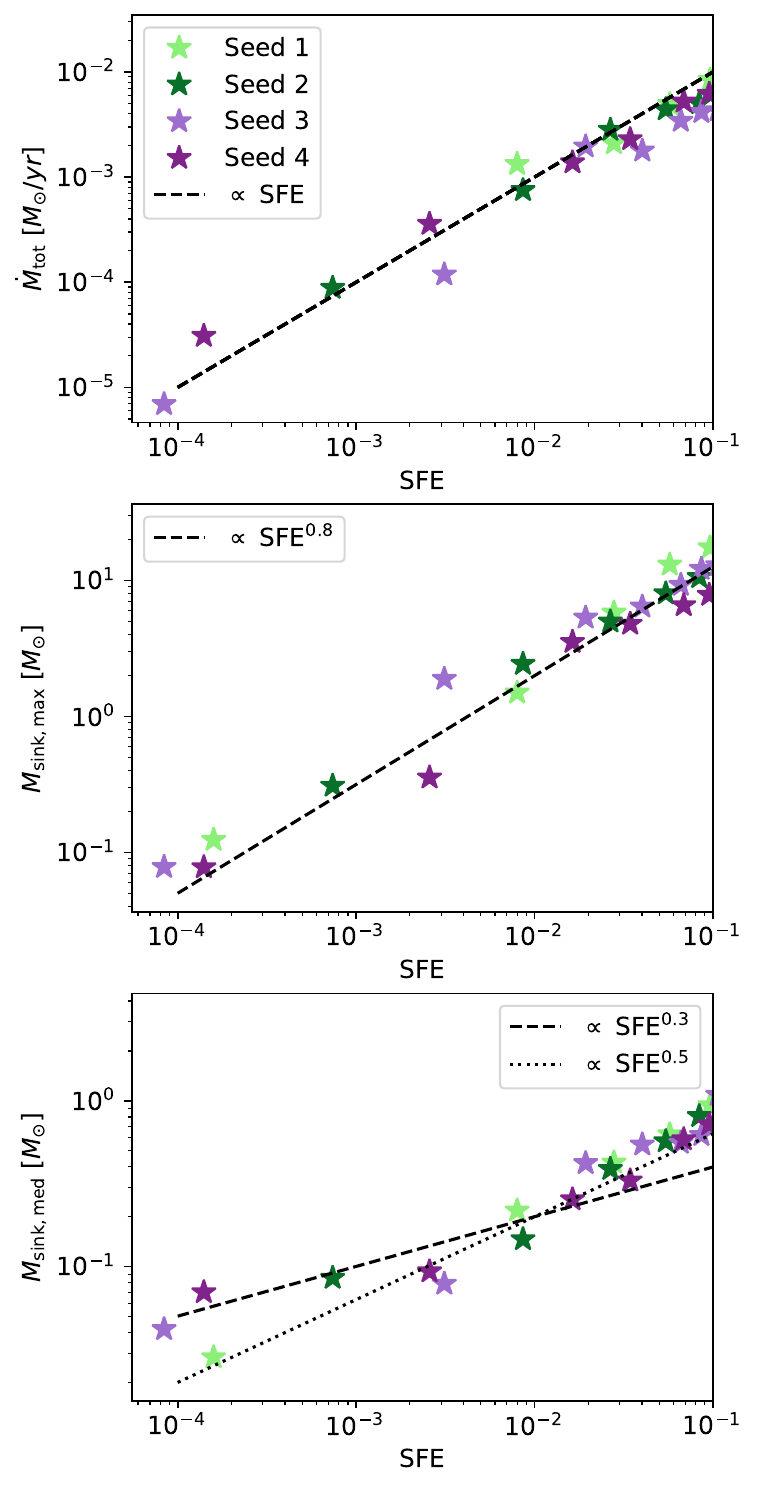}
      \caption{Evolution of the star formation for the four realizations of the reference set of parameters. From top to bottom, we show the star formation rate, the maximum sink mass, and the median sink mass (or IMF peak). We overlay some power-law correlations to guide the eye.}
            \label{fig:SF_seeds}
\end{figure}

We now describe the influence of the random initial conditions in terms of star formation. First of all, we focus on the stellar IMF. They are displayed in Fig.~\ref{fig:IMF_seeds} at SFE=0.1 for the four seeds. Despite looking very different by eye, the four realizations produce surprisingly similar IMFs, with a relatively well-defined peak in the vicinity of 1 $M_{\odot}$ and a mass range between $0.01-0.03~M_{\odot}$ and $10-20~M_{\odot}$. On top of the individual IMF, we display the average of the four clumps' IMFs. This IMF is closer to statistical convergence. However, it is still too far from convergence for us to determine whether there is a power law tail at high mass and more realizations will be needed in the future to say more on this. The peak is also more clearly defined in the case of the averaged IMF and appears to be, in fact, slightly below 1 $M_{\odot}$. 

Let us now look in more detail at the evolution of star formation as a function of SFE. Figure~\ref{fig:SF_seeds} shows the evolution of the total mass accretion rate (top), maximum sink mass (middle), and median sink mass (bottom) as a function of the SFE for the four realizations. As for the IMFs, we note that the evolution of the accretion rate and median sink mass are roughly the same for the four models \citep[see][for similar conclusion with clusters of $\sim 100 M_{\odot}$]{Girichidis2011}. The accretion rate, which increases more or less linearly with the SFE, reaches $10^{-2} M_{\odot}/$~yr at SFE=0.1. At this time, the median sink mass, which also seems to increase with SFE, is about  1 $M_{\odot}$. We note, however, that the maximum sink mass differs significantly for the four models, even if the rough scaling is similar at first glance. At SFE=0.1 it reaches  $\sim 17.6 M_{\odot}$ for the seed 1 and only $\sim 7.9 M_{\odot}$ for seed 4. This reveals the stochastic nature of high-mass star formation that does not only require a massive reservoir, but also the right conditions in terms of flow structure to happen. Interestingly, seed 4 seems to have two main sites for star formation instead of one; this could be an indication of some degree of competitive accretion \citep{1982NYASA.395..226Z,2001MNRAS.323..785B}. The randomness nature of massive star formation was also illustrated by the STARFORGE simulations \citep[e.g.][]{2021MNRAS.502.3646G}. If we now look more closely at the evolution of the median and maximum masses as a function of the SFE, we find that our findings here are very close to the predictions of STARFORGE. Indeed, as is pointed out by \cite{2021MNRAS.502.3646G}, the maximum sink mass increases with the SFE, following a scaling close to $\propto \rm{SFE}^{0.8}$, which is what we find as well. In addition, they have found that the median mass scales more or less as  $\propto \rm{SFE}^{0.3}$. In our case this seems to apply to a low SFE, but the relation might be steeper at the latter stages of the simulation (except for seed 1). Our findings here are also in line with the higher-resolution models of \cite{Hennebelleetal2022}, who noted that the maximum sink mass roughly increases linearly with the SFE.

To summarize, we have shown that except for the mass of the most massive star the choice of initial turbulent seed has a fairly small impact on the final result of the simulation when it comes to star formation and the density PDF. Interestingly, one outcome of the RS approach is that we notice that a difference emerges among these models once the full RS post-processing pipeline is applied (paper III). As is shown in paper III, at the resolution of the observations the different realizations lead to slightly different statistics. Interestingly, in paper III, we find that seed 4 is also the model that stands out the most when it comes to the number of fragments. It indeed resembles better models with $\mu=100$ than $\mu=10$ when it comes to the statistics of fragments. This must be accounted for as an uncertainty when the properties are derived from real observations. On the modeling side, an advantage of running different simulations of the same realization is that it increases our statistics for the stellar IMF and tells us more about how the most massive stars are assembled. All of this considered shows that, although the turbulent seed has a small influence on the result, it is still a key variable. This may reflect the intrinsically complex and unpredictable behavior of gravo-turbulence that is also seen in star-forming regions.

\subsubsection{Magnetic field strength}

\begin{figure*}[t!]
  \centering
 \includegraphics[width=
          0.85\textwidth]{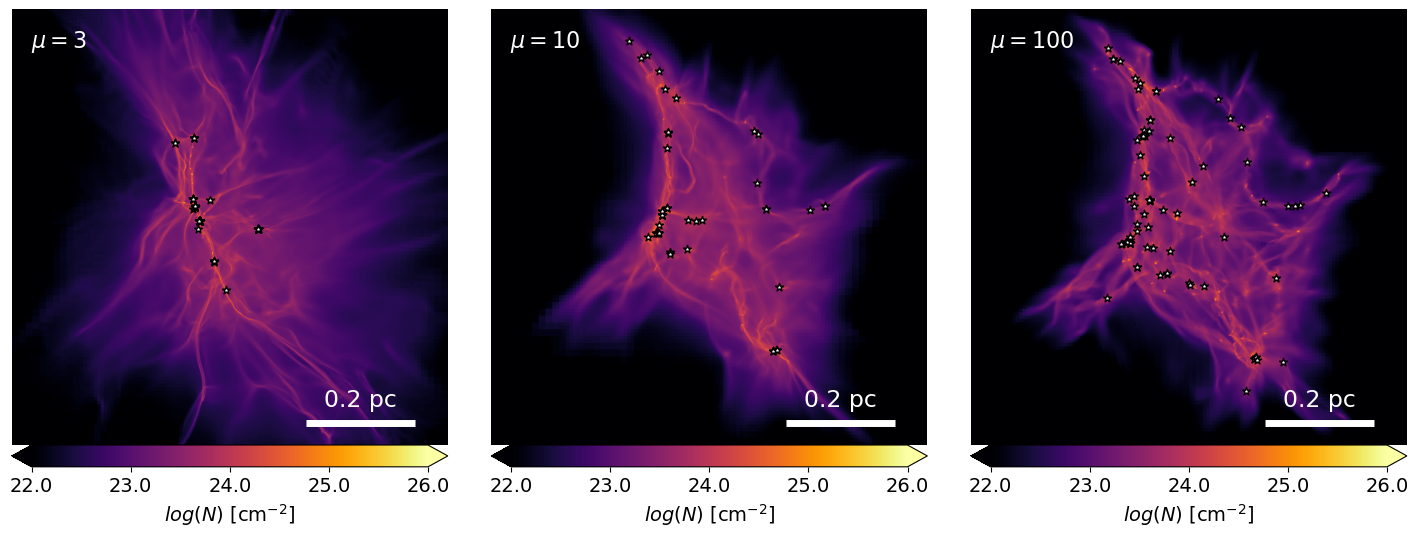}
      \caption{Illustration of the magnetic field strength impact
      on the column density (integrated in the z direction).
      The columns correspond to models with the same fiducial parameters, except for the magnetic field.
      From left to right, we display the results for mass-to-flux ratios $\mu$\,$=$\,3, 10, and 100.}\label{fig:coldens_B}
\end{figure*}

\begin{figure*}[h!]
  \centering
 \includegraphics[width=
          0.8\textwidth]{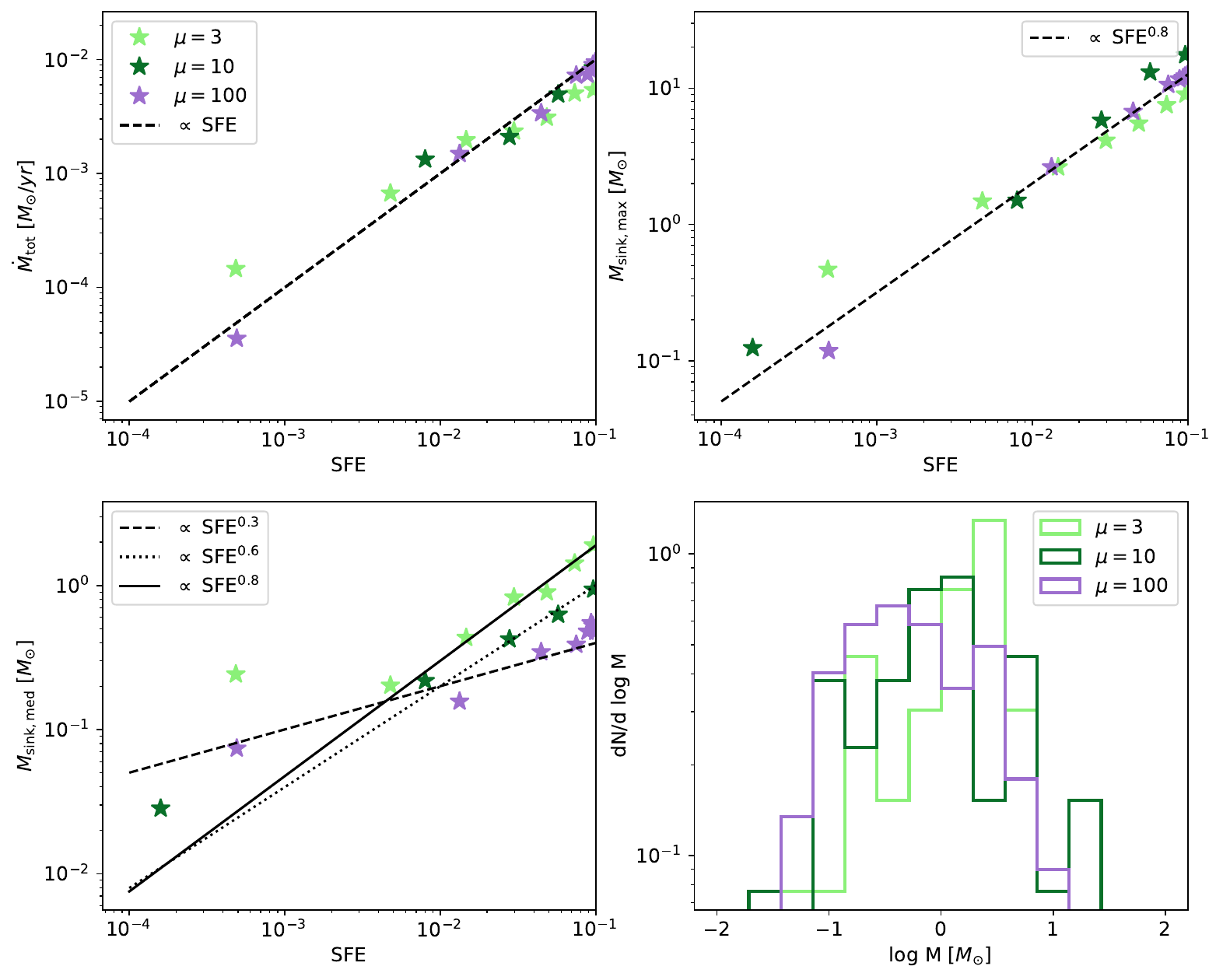}
      \caption{Evolution of star formation for the three different magnetic fields (with the fiducial set of parameters except for the magnetic field). From left to right, top to bottom, we display the star formation rate, the maximum sink mass, the median sink mass as a function of the SFE, and the IMF at SFE=0.1. We overlay some power-law correlations to guide the eye.}
            \label{fig:IMF_B}
\end{figure*}

\begin{figure*}[t!]
  \centering
 \includegraphics[width=
          \textwidth]{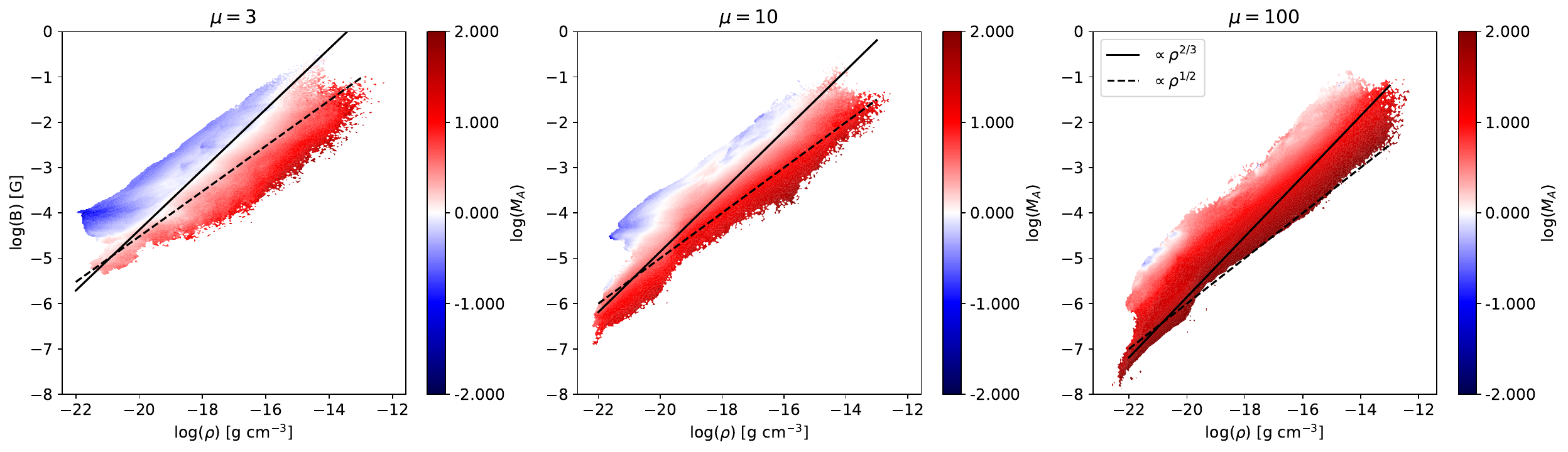}
      \caption{
      Illustration of the impact of initial magnetization on the scaling of magnetic field strength (B) with density ($\rho$). The color displays the logarithm of the Alfv\'enic Mach number.
      From left to right, we display $\mu=3$, $\mu=10$, and $\mu=100$ at SFE=0.1. We clearly see that the relation steepens toward a $\rho^{2/3}$ scaling as the $\mu$ increases ($B$ decreases). We overlay some power-law correlations to guide the eye.}
            \label{fig:B_histograms}
\end{figure*}

Depending on its strength, the magnetic field can support the clump collapse and fragmentation to various degrees. Measurements of the magnetic fields are, however, very uncertain and it is not yet clear how much they affect the star-forming clump evolution. We therefore investigated three different magnetic field intensities or, more precisely, three different mass-to-flux ratios. We considered $\mu=3$ (strong field), $\mu=10$ (moderate field), and $\mu=100$ (weak field) in order to provide a relatively complete picture of its influence. We recall that for $\mu=1$ the magnetic field is expected to completely prevent the collapse of the clump.

In Fig.~\ref{fig:coldens_B}, we show the column density, in the fiducial configuration and at SFE=0.1, for these three magnetic field strengths. Quite clearly, the magnetic field shapes the clump morphology at scales that depend on its intensity. Going from a moderate (middle panel) to a weak field (right panel), we see that the overall aspect of the clump is quite similar. This is not a complete surprise, since a mass-to-flux ratio of 10 is (by definition) already not enough to provide a significant opposition to the collapse. Of note, the small scales are quite strongly affected by the change from $\mu=10$ to $\mu=100$, the latter being clearly more fragmented. This effect was already previously noted for the higher-resolution models of \cite{Hennebelleetal2022} or \cite{Lebreuilly2024a} when going from $\mu=10$ to $\mu=50$ and is also consistent with the findings of \cite{2010A&A...510L...3C} that have shown that the combined effect of stellar feedback and magnetic pressure is very efficient at suppressing small scale fragmentation. If we now compare the  $\mu=3$ to the $\mu=10$ and  $\mu=100$ cases, the effect of magnetic field is even more dramatic, as for $\mu=3$ it visibly changes the large-scale evolution of the clump. With  $\mu=3$, the clump is much more organized as sheets in the magnetic field direction (z). We also note that, as the magnetic field increases from $\mu=100$ to $\mu=3$, fragmentation is not totally suppressed but happens more and more locally in the clump. While the distribution of cores is more or less homogeneous in the $\mu=10$ case, they are almost exclusively in a single filament in the $\mu=3$ case. Almost everywhere else in the clump, star formation is precluded when the magnetic field is strong. Quite naturally, the number of sinks drastically diminishes with the mass-to-flux ratio, which goes from 78 to 46. Of course, this lack of fragmentation also has dramatic consequences for the stellar mass spectrum evolution, as can clearly be seen in the Fig.~\ref{fig:IMF_B}. While the star formation rate of the three models is similar, we clearly see that the more magnetized model promotes the formation of a larger number of stars above $1 M_{\odot}$, with a median stellar mass increasing more steeply with the SFE for $\mu=3$. By SFE=0.1, this median reaches almost $3 M_{\odot}$ while it is around $0.8 M_{\odot}$ for the $\mu=100$ case. Quite counterintuitively, the mass of the most massive star seems, however, to be decreasing with an increasing magnetic field (or at least going from $\mu=10$ to $\mu=3$). This might be a consequence of the increased significant of the magnetic support against the global collapse, which prevents the formation of a main star-forming hub that is continuously fed by the network of filaments in the low magnetic field cases, or simply of insufficient statistics.

As a complement to this description, we show in Fig.~\ref{fig:B_histograms} the $B-\rho$ relation for the three models with increasing $\mu$ from left to right. We also display in color the logarithm of the Alfv\'enic Mach number. As can be seen, in all three models B increases with $\rho$; however, the $B-\rho$ changes when going from $\mu = 3$, where $B \propto \rho^{1/2}$, to $\mu=100$, where the scaling  $B \propto \rho^{2/3}$ is quite clear. We clearly see that the flow is in large part sub-Alfv\'enic for $\mu = 3$, and completely super-Alfv\'enic for $\mu=100$. The transition between the two scalings seems to happen between $\mu=10$ and $\mu =3$. As was explained earlier, the $\mu=10$ case is intermediate. However, if the two behaviors are observed in a different part of the clump, $B \propto \rho^{2/3}$ seems to be the dominant scaling, which makes sense considering that most of the flow is super-Alfv\'enic.  This variety in the $B-\rho$ between our different cases shows that this scaling might not be universal if there is a diversity in clump magnetization.

\subsection{Lower-mass clumps and higher Mach numbers}

\begin{figure*}[h!]
  \centering
 \includegraphics[width=
          0.8\textwidth]{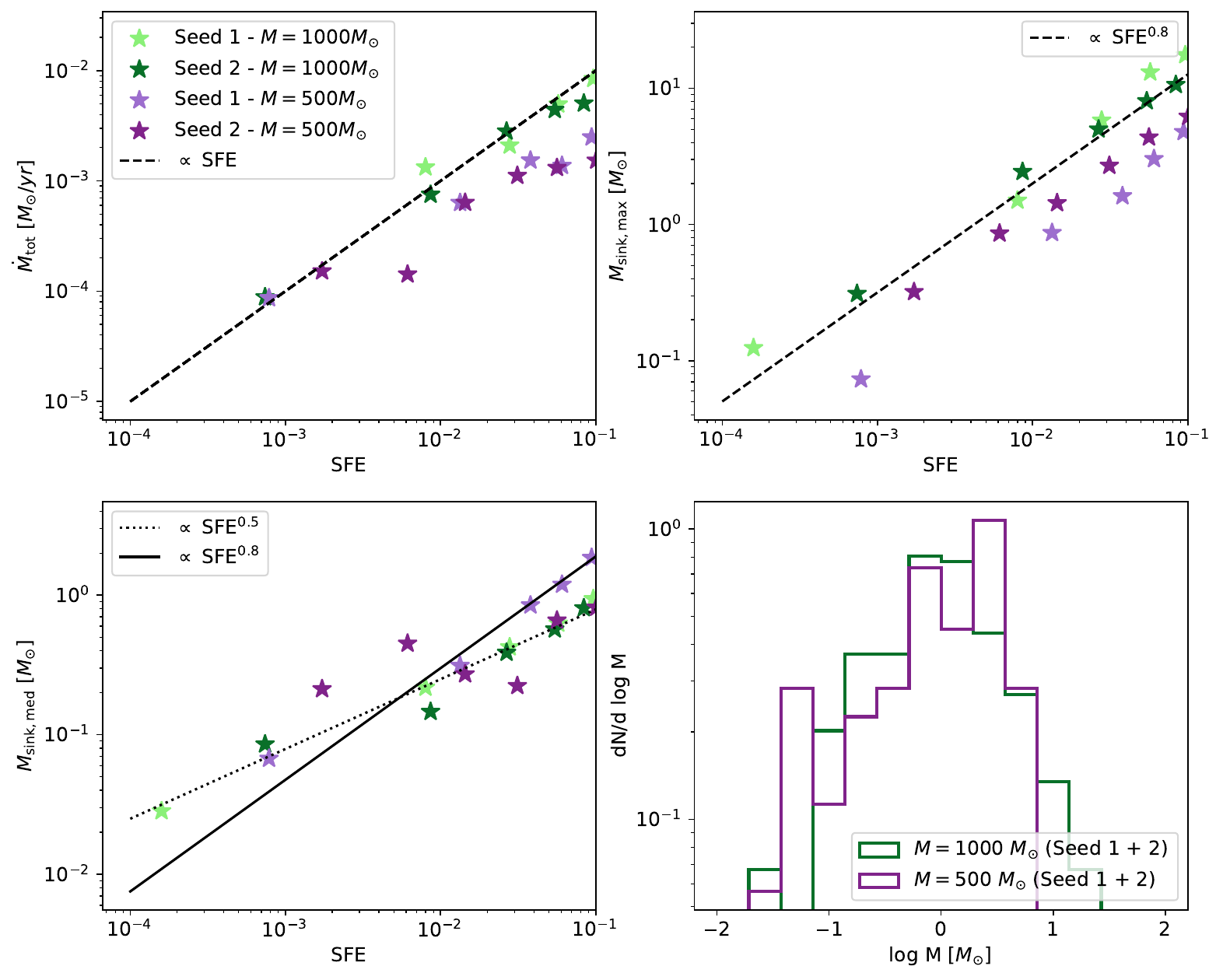}
      \caption{Evolution of star formation for the clumps of 1000 and 500 $M_{\odot}$ for seeds 1 and 2 (with the fiducial set of parameters, except for the mass). From left to right, we display the star formation rate, the maximum sink mass, the median sink mass as a function of the SFE, and the IMF at SFE=0.1. For the IMF, we combined seed 1 and 2 to enhance the statistics of the 500 $M_{\odot}$ case. }
            \label{fig:IMF_mass}
\end{figure*}
\begin{figure*}[h!]
  \centering
 \includegraphics[width=
          0.8\textwidth]{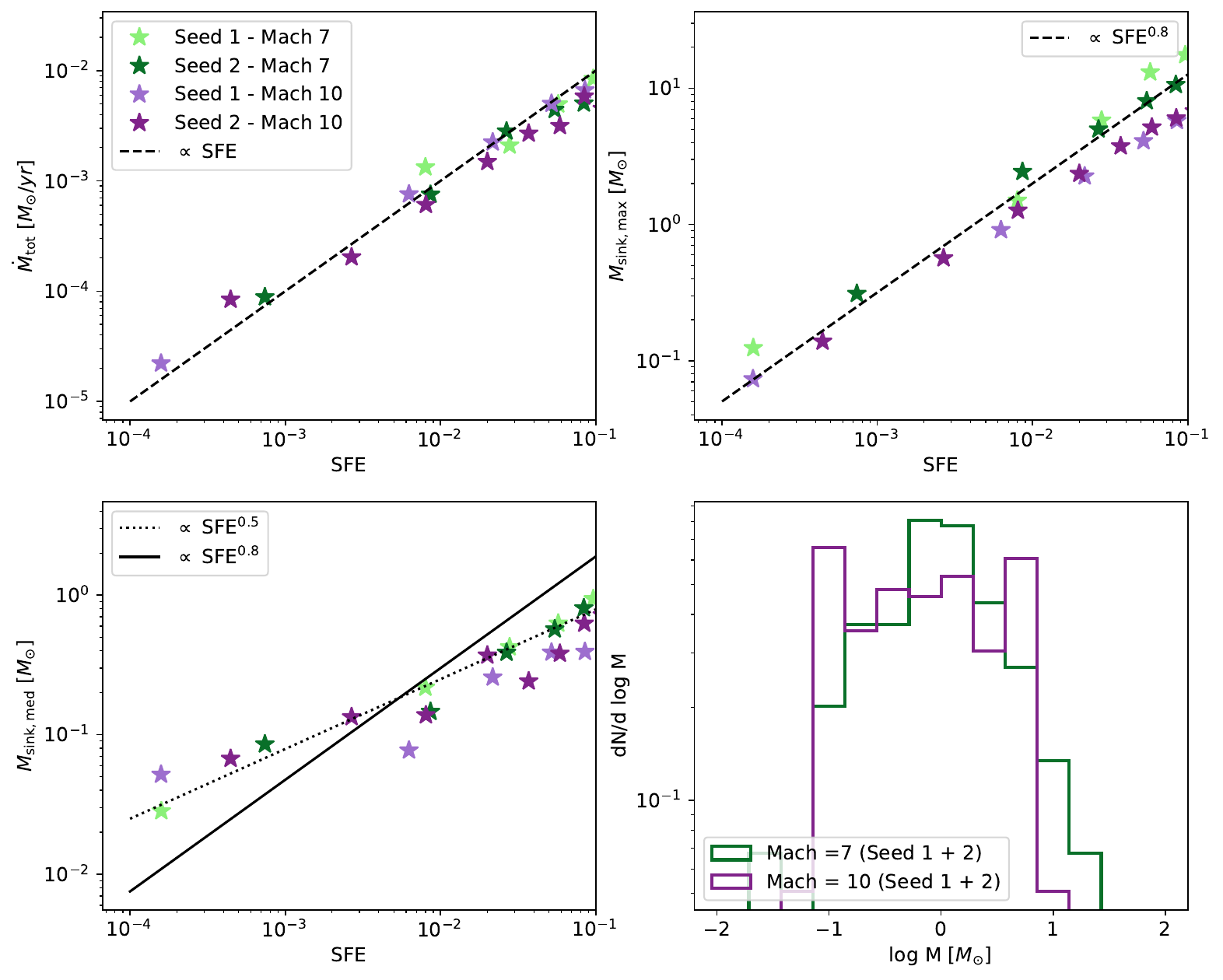}
      \caption{Same as Fig~\ref{fig:IMF_mass} but varying the Mach number instead of the mass. }
            \label{fig:IMF_mach}
\end{figure*}

Finally, we focus on the influence of the clump surface density, by comparing our reference case with a run of $500 M_{\odot}$. We also consider the lesser influence of the Mach number\footnote{in our narrow range chosen to reproduce the SQUALO conditions}, by comparing the fiducial model to a run with $\mathcal{M}=10$. We focus here on star formation because the clump structure is not strongly affected by these two parameters (except of course the normalization for the column density). In Fig.~\ref{fig:IMF_mass} and Fig.~\ref{fig:IMF_mach}, we show the influence of these two parameters on the star formation rate, the maximum and median sink mass, and the stellar IMFs (at SFE=0.1). We note that in both cases we combined the data of seeds 1 and 2 to get more statistically significant samples for the IMFs. We see that reducing the clump mass lead to a reduced star formation rate, particularly at a high SFE. This is not surprising, since the 500 $M_{\odot}$ clump is more diffused and more stable against collapse (as we kept the same radius for the two clumps). Changing the Mach number also reduces the SFR but not as strongly. Similarly and also unsurprisingly, the maximum sink mass is also reduced by the process of decreasing the clump mass or increasing the Mach number. This correlation between the cluster mass and the mass of the most massive star goes in favor of a more clump-fed scenario \cite{2004MNRAS.349..735B}.  The influence of the clump mass on the median sink mass is not as obvious, although it also seems to be slightly increasing with a decreasing clump mass (at least for seed 1). Although we see that the median mass is slightly reduced when increasing the Mach number, we can clearly see that its strongest impact is on the stellar IMF, which plateaus more significantly for the Mach 10 case. This is at odds with the previous finding of \cite{Hennebelleetal2022}, in which the opposed effect was found. This might be a consequence of the low resolution used here. Clearly we do not resolve the first Larson cores, and therefore an excess of low-mass star formation could be considered to be artificial.

\section{Discussion}
\label{sec:discussion}

\subsection{Whether $L/M$ is a good evolutionary tracer for clumps}

\begin{figure*}[t!]
  \centering
 \includegraphics[width=
         0.8\textwidth]{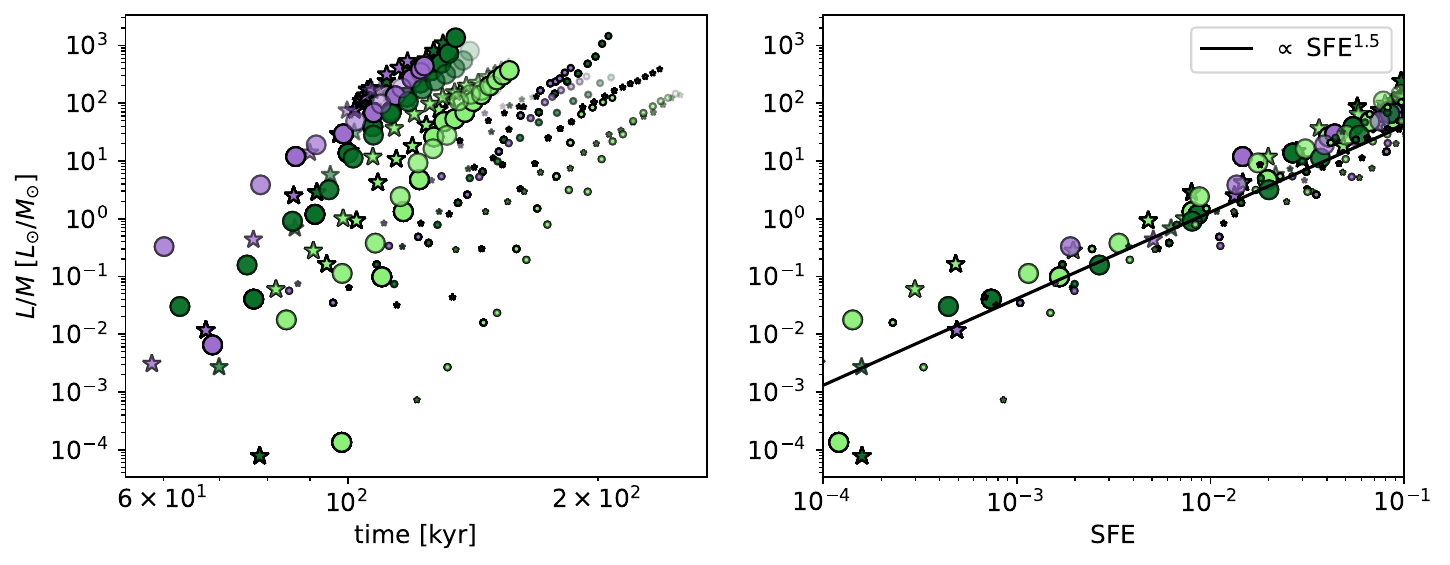}
      \caption{L/M as a function of the time (left) and the SFE (right) for the full RS1.0 catalog of models. Here large (resp. small) markers represent 1000  (resp. 500)  $M_{\odot}$clumps. We use three different colors to represent the different mass-to-flux ratios (purple is 100, dark green is 10, and light green is 3). In addition, the circles represent seed 1, while the star symbols represent seed 2. Finally, we use the transparency of the markers to display the two Mach numbers (plain markers for Mach 7 and transparent markers for Mach 10).}
            \label{fig:LoverM}
\end{figure*}

\begin{figure*}[t!]
  \centering
 \includegraphics[width=
         0.8\textwidth]{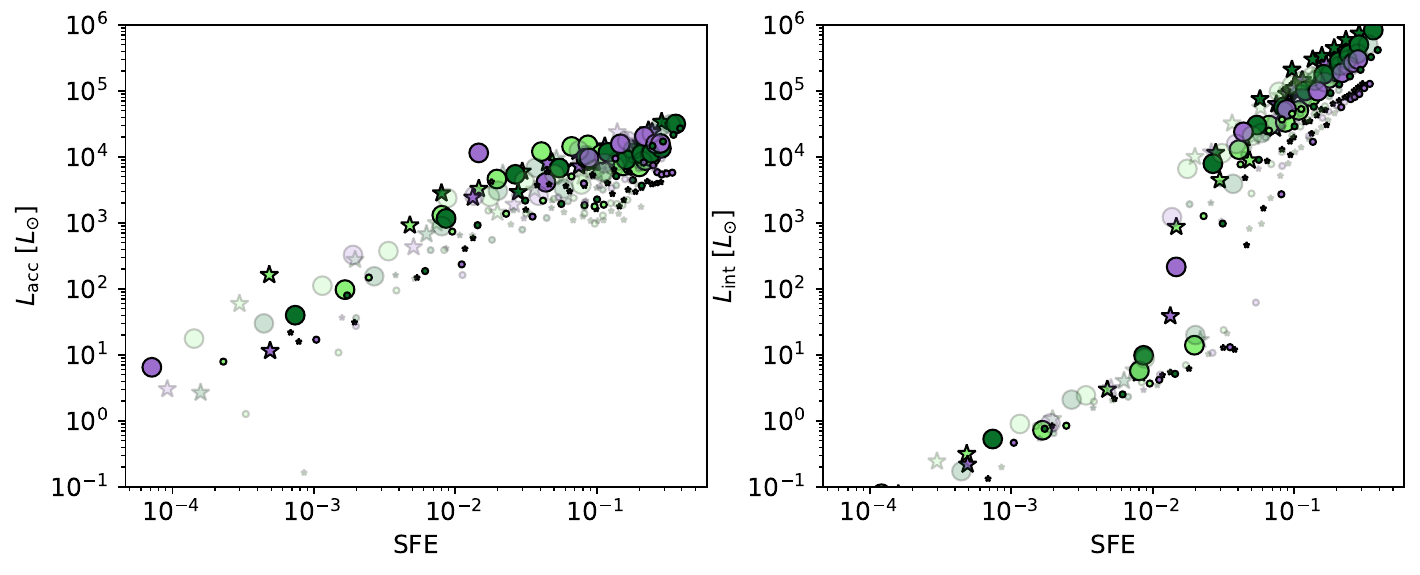}
      \caption{$L_{\mathrm{acc}}$ (left) and $L_{\mathrm{int}}$ (right)   as a function of the time (left) and the SFE (right) for the full RS1.0 catalog of models. The color and marker coding is the same as in Fig.~\ref{fig:LoverM}.}
            \label{fig:LaccvsLint}
\end{figure*}

The $L/M$ is a widely used diagnosis to quantify the evolutionary stage of real star-forming clump. In this section, we discuss the viability of this tool in the context of our RS1.0 models. We recall that the value of $L/M$ measured here is not the final product of the post-processing routine that we shall discuss in Paper II but rather an intermediate state directly measured from the simulation. The luminosity is simply the total sink luminosity and the mass is the remaining gas mass, i.e., the total clump mass minus the sink mass. As such this $L/M$ measurement does not account for the reprocessing of the stellar light by the clump or the uncertainties of the mass measurement when converting a flux to a mass. 

We show in Fig~\ref{fig:LoverM} the $L/M$ extracted for all the models of the RS1.0 catalog as a function of the time on the left and SFE on the right. For these two plots we follow a very specific scheme: large (resp. small) markers represent 1000  (resp. 500)  $M_{\odot}$ clumps, the three different colors represent the three mass-to-flux ratios (purple is 100, dark green is 10, and light green is 3), the circles represent seed 1, while the star symbols represent seed 2, and finally the plain markers represent Mach 7, while the transparent ones represent Mach 10. First of all, it is quite clear that $L/M$ increases almost monotonically as the model evolves (either with time or SFE). It goes from very low values of $\sim 10^{-4}-10^{-3} L_{\odot}/M_{\odot}$ at early times up to values on the order of  $\sim 10^{3} L_{\odot}/M_{\odot}$ at the final stages of the simulations. This is indicates that $L/M$ is indeed a clear marker of clump evolution. However, quite evidently it cannot be used as a measure of time. As can be seen in the left panels, at a given $L/M$ it is impossible to recover the time without any information on the model. This makes sense because simulations with different masses, magnetizations, or Mach numbers are expected to have different collapse timescales, and therefore to start forming stars at very different times. For example, we can clearly see that the 1000 $M_{\odot}$ clumps (large markers) achieve a high $L/M$ faster than the 500 $M_{\odot}$ clumps (small markers). Within groups of models of similar mass, we see that the magnetic field is also very influential. The evolution of $\mu=3$ clumps (light green) is quite evidently slower for $\mu=10$ and $\mu=100$ (darker green and purple) because the magnetic field is providing some additional support to the collapse. 

Strikingly, this wide spread disappears when displaying $L/M$ as a function of the SFE. At a given SFE, there is only a uncertainty of $\sim 1$ dex $L/M$. This gives hope that $L/M$ can be used as an evolution tracer for real star-forming clump, provided that this relation stands still after full post-processing of the simulation (see Paper II, in which we show that this is actually the case).  It is interesting, then, to determine where this relation comes from. We display in Fig.~\ref{fig:LaccvsLint} the evolution of the total accretion (left) and internal (right) luminosity as a function of the SFE. From these two plots, it becomes evident that the previous scaling is, in fact, the combination of two very similar scalings. At SFE<0.01, the accretion luminosity largely dominates the internal luminosity. Above this the internal luminosity sharply rises and begins to dominate the total luminosity budget. This happens because stars larger than $2~M_{\odot}$ start to form. At SFE>0.01 the scaling between the accretion luminosity and the SFE becomes shallower. 

Let us now try to understand the relation between $L/M$ and the SFE. For this, we keep in mind that we have previously shown that $\dot{M} \propto \mathrm{SFE}$ and that the typical relation between the median mass and the SFE in our models is on the order of $\propto \mathrm{SFE}^{\sim 0.5}$. For simplicity, we assume that all stars in the clump are identical and have a mass on the order of the median sink mass at early stages and that the dominant source of luminosity is the most massive star at later stages. We recall that the denominator, M, of the $L/M$ ratio corresponds to the remaining clump mass, which we assume to be the initial one as we only look at SFE<0.1. 

For a clump with only low-mass stars (when  $M_{\mathrm{max}}<2 M_{\odot}$), the accretion luminosity dominates and we can therefore write 

\begin{equation}
    L/M \propto \frac{\dot{M}M_{\mathrm{med}}}{R}.
\end{equation}
From the tables of \cite{2013ApJ...772...61K}, we see that for low-mass objects the stellar radius is typically constant and on the order of the solar radii. We therefore determine that
\begin{equation}
    L/M \propto \mathrm{SFE}^{1.5}
.\end{equation}
This is typically what happens at early stages during the clump evolution. When the internal luminosity starts to dominate, the scaling changes slightly as $L_{\mathrm{int}} \propto M_{\mathrm{max}}^2$ (when it is $>2 M_{\odot}$), which results in 
\begin{equation}
    L/M \propto \mathrm{SFE}^{1.6}
,\end{equation}
since we typically have $M_{\mathrm{max}} \propto \mathrm{SFE}^{0.8}$. This scaling is very close to the scaling at low mass so we can assume they are the same (although the origin is different).
As can be seen in Fig~\ref{fig:LoverM}, these scalings are in relatively good agreement with what we observe in our models (with some spread). Of course, we can now wonder if this law is general. Indeed, one of the key physical ingredients that we did not consider here is the occurrence of protostellar jets. Jet components are ubiquitous for clumps, whatever the value of $L/M$ \citep{2019ApJ...886...36S}. Because our resolution is on the order of 40 au, jets do not occur self-consistently in the model. Therefore, they should be included as a sub-grid modeling. While jets are indeed implemented in RAMSES \citep{2022A&A...663A...6V}, we did not consider them here. The main reason of this choice is to reduce the numerical cost of running many models. Jets indeed provoke very high velocities, which substantially decreases the timestep. Nevertheless,  they have been found to play an important role, especially for the formation of the most massive stars, and to reduce the SFR \citep{2015MNRAS.450.4035F}. It has also been shown that they are influential for the IMF. \cite{2024A&A...683A..13L} have shown that the shape of the IMF is significantly influenced by the presence of jets in our fiducial set of parameters, due to their combined influence on the dynamics of the clump through the kinetic energy they inject and on its thermal evolution by reducing the stellar accretion luminosity. Other works even argued that the jets are critical for setting the peak of the IMF \citep{2010ApJ...720L..26L,2021MNRAS.502.3646G,2021MNRAS.507.2448M}. Given their importance, the next set of Rosetta Stone simulations will include the jets despite their numerical price.

    Another prominent feedback effect that could play a critical role in the dissipation of the clump, and therefore the regulation of star formation, is the expansion of HII regions. This effect is also implemented in RAMSES but only when using a M1 treatment for the radiation. It is therefore incompatible with the FLD radiative transfer that we use here to efficiently treat the infrared photons. HII regions are prominent in observations for $L/M>10$ \citep{2015A&A...579A..71C}. As can be seen, this happens in our models at a SFE on the order of a few 0.01. This is completely consistent with the steep rise of the internal luminosity in our models, shown in Fig.~\ref{fig:LaccvsLint}. HII regions are widely believed to play an important role in the evolution of star-forming clumps, especially in their destruction and the generation of Pillars of Creation-like structures hosting the next generation of protostars \citep[see e.g.][]{2011MNRAS.414..321D,2013MNRAS.435..917W,2014MNRAS.442..694D,2015MNRAS.454.4484G,2022A&A...663A...6V,2024A&A...682A..76S}. As such,  {he future RS models that will be computed will require special attention to be paid to the radiative transfer, including the impact of HII regions.

 All things considered, it is important to keep in mind that the relation between L/M and  SFE is not perfect and probably more complex in real clumps than the one presented here. Hence, this intermediate result needs to be taken with a grain of salt.

\subsection{Effect of the sink threshold and numerical resolution}

\begin{figure*}[t!]
  \centering
 \includegraphics[width=
          0.85\textwidth]{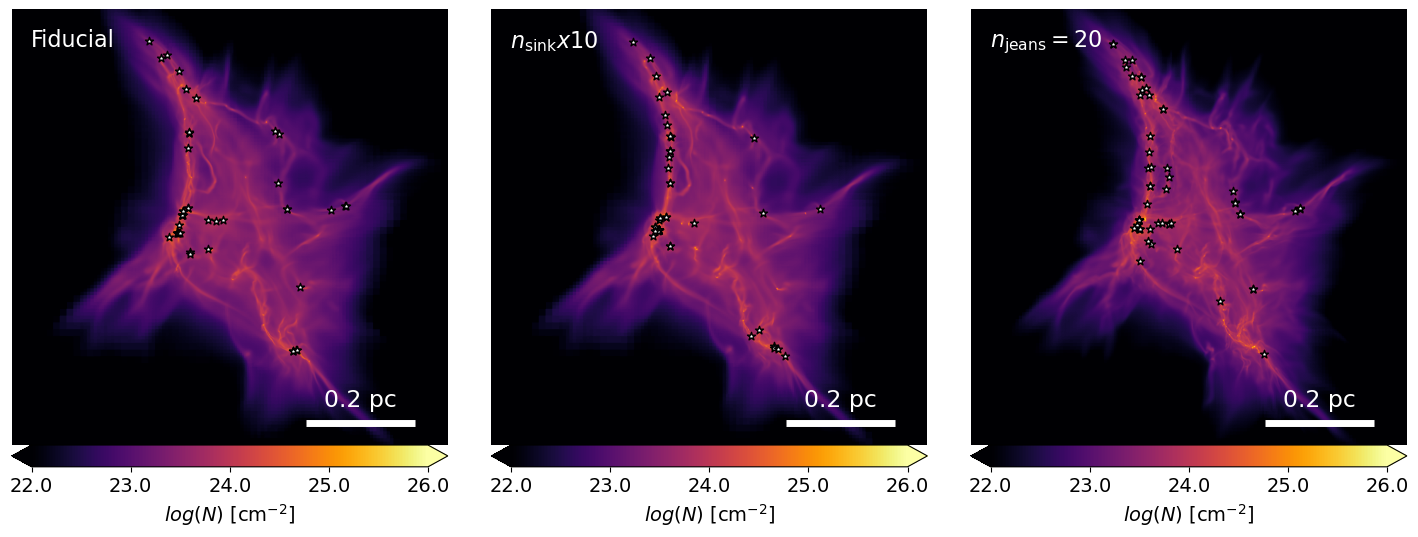}
      \caption{
      Impact of the choice of sink threshold and resolution on the column density (integrated along the $z$ axis) and sink formation.
      }\label{fig:coldens_num}
\end{figure*}

The issue of finite resolution can clearly be a problem in numerical simulations when some important processes happen at scales that are not resolved by models. For the Rosetta Stone project, we made the deliberate choice of limiting the maximal resolution to $\sim 38$~au to run a large set of simulations. Contrary to our previous works \citep{Lebreuilly2021}, we therefore cannot describe the opacity limit properly. As has been noted by several studies, this transition between the isothermal and adiabatic regime could play a fundamental role in setting the peak of the IMF \citep{2019ApJ...883..140H,2020MNRAS.492.4727C,2023arXiv230816268G,IMFpatrick}. In addition, our resolution does not allow us to resolve protostellar disks and their potential fragmentation, which could (and does) lead to the formation of tight multiples. While we mitigate the effect of low resolution by setting a sink creation threshold, which allows us to recover an IMF peak at a position similar to the one obtained in more resolved models, we emphasize that this peak is still artificial and is not guaranteed to be correct for all the models in our catalog.  It is also important to recall that the resolution at intermediate scales is also very important. As was explained earlier, we ran our catalog of simulations using 10 points per Jeans length, which allowed us to verify the criterion of \cite{1997ApJ...489L.179T} to avoid artificial fragmentation. But this criterion might be insufficient in the case of a gravo-turbulent collapse or in the presence of magnetic fields \citep{2011ApJ...731...62F}. While it is unfortunately unfeasible to run the full suite of simulations with a higher resolution, we still explore our reference configuration with 20 points per Jeans length.

We show in Fig.~\ref{fig:coldens_num} the column density integrated along the z axis for this model as well as the fiducial model. In addition, we display a model run for which we varied the sink threshold, imposing $n_{\mathrm{sink}}=10^{10} \centi\meter^{-3}$. Let us first focus on the impact of resolution. Quite clearly, the clump with $N_{\mathrm{jeans}}=20$ shows more structures at the small scale, which can be explained as a decrease in diffusivity with the increase in resolution. Of course, this comes with a heavy numerical cost that cannot be afforded for a full catalog of simulations. This justifies why we still go for $N_{\mathrm{jeans}}=10$.  It is also important to quantify the uncertainty that is brought by our choice of sink threshold. The variations between $n_{\mathrm{sink}}=10^{10} \centi\meter^{-3}$ and $n_{\mathrm{sink}}=10^{9} \centi\meter^{-3}$ seem to be quite minor. Nevertheless, some small differences can be observed. While star formation mostly proceeds at the same locations in the two models, there are more stars in the upper left filament in the $n_{\mathrm{sink}}=10^{10} \centi\meter^{-3}$ case. If we now focus on the case with $N_{\mathrm{jeans}}=20$, we find a sink distribution that agrees more with the $n_{\mathrm{sink}}=10^{10} \centi\meter^{-3}$ case than with our fiducial model. While this is surprising at first, this could be due to the small delay in the star formation process that is brought by the slightly better resolved turbulence. Small-scale structures remains stable for a slightly longer time and as a result the clump has more time to fragment before forming stars. At this stage it is useful to recall that formed stars influence the fragmentation process through their luminosity (internal or accretion). This is why the exact moment at which they form, and the synchronization of the star formation process, matter. To summarize, while the numerical effects appear to have a limited influence on our results, it is important to recall that several differences exist.

\section{Conclusion}
\label{sec:conclusion}

In this article we have presented the first catalog of models of the Rosetta Stone project RS1.0. We have explored a full grid of star-forming clump models computed with the MHD AMR code RAMSES assuming two different initial masses, three different magnetizations  ($\mu$ = 3, 10, 100), two different Mach numbers (7,10), and two different realizations (four in a specific case). As a complement and to help our discussion, we also computed and presented four additional models. We summarize our various findings here.

\begin{itemize}
    \item We present the first stone of our Rosetta Stone project. We recall that the aim of the project is to provide the community with a complete end-to-end framework for comparing high-mass star-forming clump models to their observed counterpart from parsec scales down to prestellar core scales.
    \item Our clump calculations all undergo gravitational collapse in a non-isotropic way as a result of turbulence and magnetization. The clump profile and temperature structure is consistent with globally collapsing clumps. These clumps are stratified in density, temperature, and magnetization.
    \item The relation between the density and magnetic field strength is either $ B\propto \rho^{1/2}$ for strong magnetic fields ($\mu=3$) or $ B\propto \rho^{2/3}$ for weaker magnetic fields ($\mu=100$). Intermediate magnetization shows an intermediate B-$\rho$ relation.
    \item The initial seed choice for the turbulent conditions mostly affects the mass of the most massive star in the clump. Other properties such as the density PDF and the peak of the IMF seem to be poorly influenced by the choice of random seed. Nevertheless, the four different seeds produce four visually different clumps and can be combined to produce a more robust IMF.
    \item The Mach number and clump initial surface density also affect star formation, with a lower Mach number and higher surface density promoting the formation of more massive stars. 
    \item We show that $L/M$ is a good measure of the clump evolution that correlates well with the SFE regardless of our choice of initial condition. We point out that the SFE versus $L/M$ relation is relatively independent of our model. However, $L/M$ is not a great measure of time, since clumps of different properties do not form stars at the same rate. 
    \item Although there seems to be a single scaling for L/M versus the SFE, there are actually two regimes in that relation. At a low SFE, the luminosity is dominated by the total accretion luminosity, while the total internal luminosity dominates later on. 
\end{itemize}

To finish, we stress that this particular work is only the first part of our complete end-to-end framework, which has a deeper meaning when considering Papers I, II, and III as one. In Paper II, the models are post-processed with a radiative transfer code at Herschel and ALMA wavelengths in order to produce synthetic observations. In this paper, we also discuss the $L/M$ relation after full post-processing. Finally, Paper III uses ideal sky maps to produce synthetic observations that are then compared to the SQUALO survey. This allows us to find a future direction in which to develop our models, guiding us in interpreting observations and devising future surveys. Future Rosetta Stone catalogs of simulations with larger parameter spaces surveyed and more physical effects (especially jets, HII regions) and scales (kiloparsecs and protostellar disks) included will be run in order to make progress on the questions related to star formation at all scales.

 \begin{acknowledgements}

We thank the referee Christoph Federrath for helping us improve the quality of this manuscript. This project was funded by the European Research Council via the ERC Synergy Grant ``ECOGAL'' (project ID 855130). We thank the whole consortium for the stimulating discussions which helped us tremendously through this process. 
AT gratefully acknowledges support from a mini-grant funded
by INAF. We acknowledge PRACE for awarding us access to the JUWELS supercomputer. This work was also granted access to HPC resources of CINES and CCRT under the allocation A0130407023 made by GENCI (Grand Equipement National de Calcul Intensif). Upon publication, the simulations of the project will be made available at \url{http://www.galactica-simulations.eu/db/STAR_FORM/ROSETTA/}.
 
\end{acknowledgements}
\bibliographystyle{aa}
\bibliography{ref}
\appendix
\onecolumn
\section{Evolution of the SFE and number of sinks}

\begin{figure*}[h!]
  \centering
 \includegraphics[width=
         0.8\textwidth]{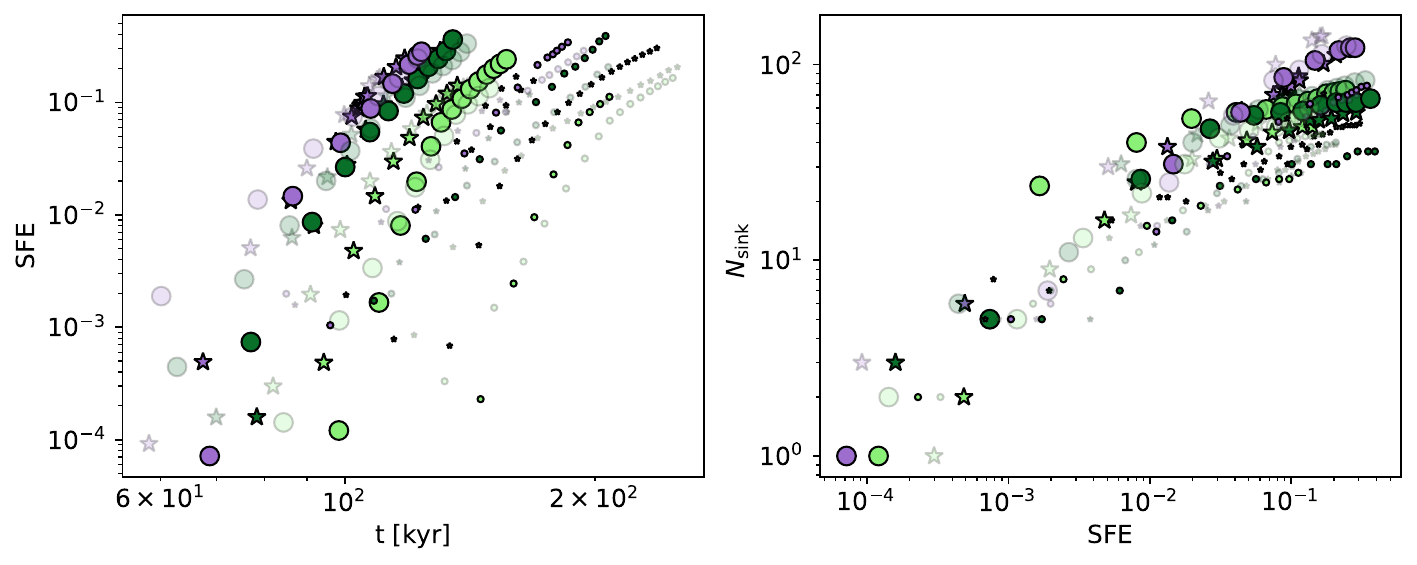}
      \caption{ SFE as a function of time (left) and number of sinks as a function of SFE (right) for all the RS1.0 models. Here large (resp. small) markers represent 1000  (resp. 500)  $M_{\odot}$clumps. We use three different colors to represent the different mass-to-flux ratios (purple is 100, dark green is 10, light green is 3). In addition, the circles represent seed 1 while the star symbols represent seed 2. Finally, we use the transparency of the markers to display the two Mach numbers (plain markers for Mach 7 and transparent markers for Mach 10).}
            \label{fig:NSINK}
\end{figure*}
In Fig.~\ref{fig:NSINK} we show the evolution of the SFE as a function of time (left) and of the sink number as a function of the SFE (right) for all our models. We follow the same color scheme as in Fig~\ref{fig:LoverM}, large (resp. small) markers represent 1000  (resp. 500)  $M_{\odot}$ clumps, the three different colors represent the three mass-to-flux ratios (purple is 100, dark green is 10, light green is 3), the circle represent seed 1 while the star symbols represent seed 2, finally the plain markers have Mach 7 while the transparent ones have Mach 10. Unexpectedly, the SFE rises with time for all the models, but we clearly see 1) that an increasing magnetic field strength (green colors) delays star formation 2) that a lower initial clump surface density (smaller markers) also delays it. Depending on the models conditions, we typically reach a SFE of 0.1 between 100 and 200 kyr, which is of the order of the free-fall timescale which is $\sim 120$~kyr for the $1000 M_{\odot}$ clumps and about $\sim 170$~kyr for the  $500 M_{\odot}$  clumps. As can be seen on the right panels, each model forms between a few tenths and up to about 100 sinks. Again, the poorly (resp. strongly) magnetized models generally form more (resp. less) sinks, which is fully consistent with an expected increased (resp. decreased) fragmentation because of a lack (resp. excess) of magnetic pressure. 

\end{document}